\documentclass[5p]{elsarticle}
\usepackage{lineno}
\usepackage{graphicx}
\usepackage[breaklinks=true]{hyperref}
\usepackage{booktabs}
\usepackage[utf8]{inputenc}
\usepackage{multirow}
\usepackage{array}
\usepackage{listings}
\usepackage{pdflscape}
\usepackage{amsmath}
\usepackage[caption=false]{subfig}
\usepackage{todonotes}
\usepackage{xcolor,colortbl}
\usepackage{adjustbox}
\usepackage[linesnumbered,ruled,vlined]{algorithm2e}

\newcolumntype{x}[1]{>{\centering\arraybackslash\hspace{0pt}}p{#1}}
\newcolumntype{C}{>{\centering\arraybackslash}m{2cm}}

\newcommand{\f}{\mkern-2mu f\mkern-3mu}

\makeatletter
\lst@InstallKeywords k{attributes}{attributestyle}\slshape{attributestyle}{}ld
\makeatother

\definecolor{bluekeywords}{rgb}{0,0,1}
\definecolor{greencomments}{rgb}{0,0.5,0}
\definecolor{redstrings}{rgb}{0.64,0.08,0.08}
\definecolor{xmlcomments}{rgb}{0.5,0.5,0.5}
\definecolor{types}{rgb}{0.17,0.57,0.68}
\begin{document}

\begin{frontmatter}

\title{Automated Isolation for White-box Test Generation}

\author{David Honfi}
\ead{honfi@mit.bme.hu}
\author{Zoltan Micskei}
\ead{micskeiz@mit.bme.hu}
\address{Department of Measurement and Information Systems,\\Budapest University of Technology and Economics,\\Budapest, Hungary}

\begin{abstract}
    \textbf{Context.} White-box test generation is a technique used for automatically selecting test inputs using only the source or binary code. However, such techniques encounter challenges when applying them to complex programs. One of the main challenges is handling the dependencies of the unit under test.

    \noindent\textbf{Objective.} Without proper actions, generated tests cannot cover all parts of the source code, or calling the dependencies may cause unexpected side effects (e.g., file system or network access). These issues should be tackled while maintaining the advantages of white-box test generation.

    \noindent\textbf{Method.} In this paper, we present an automated source code transformation approach tackling the dependency issue for white-box test generation. This technique isolates the test execution by creating a parameterized sandbox wrapped around the transformed unit. We implemented the approach in a ready-to-use tool using Microsoft Pex as a test generator, and evaluated it on 10 open-source projects from GitHub having more than 38.000 lines of code in total.

    \noindent\textbf{Results.} The results from the evaluation indicate that if the lack of isolation hinders white-box test generation, then our approach is able to help: it increases the code coverage reached by the automatically generated test, while it reduces unwanted side effects. Also, our results act as a unique baseline for the test generation performance of Microsoft Pex on open-source projects.

    \noindent\textbf{Conclusion.} Based on the results, our source code transformations might serve well for alleviating the isolation problem in white-box test generation as it increases the coverage reached in such situations, while maintaining the practical applicability of the tests generated on the isolated code.
    
\end{abstract}



\end{frontmatter}


\section{Introduction}\label{sec:intro}

\subsection{Context and motivation}

In software engineering, testing is one of the most frequently used techniques to enhance the quality of software. Developing test usually requires a significant amount of time due to its complexity. To improve testing, several automated techniques have already been proposed. A subset of these techniques use the source or binary code (i.e., the implementation) to generate test inputs and observe their outputs, which are commonly called as \emph{white-box} tests. These techniques include symbolic execution \cite{classic-se}, dynamic symbolic execution (also known as concolic testing) \cite{concolic}, search-based test generation \cite{sbst}, and random test generation \cite{random}.

In the recent years these methods have been implemented in ready-to-use tools -- such as Pex (IntelliTest) \cite{pexwb}, Randoop \cite{randoop}, KLEE \cite{klee}, or EvoSuite \cite{evosuite} -- to enhance unit level-testing. Although these may help reducing effort and time spent on testing, they commonly encounter challenges in complex software that hinder their execution \cite{sbst,random,se-decades}. From these challenges, our paper focuses on \emph{handling environment dependencies}: if the code under test heavily depends on its environment (e.g., file system, network, or other modules), it may negatively affect test results. These dependencies can cause unexpected behavior, flaky tests or infeasible code exploration. To avoid such issues, the dependencies should be isolated for testing; this is commonly referred to as the \emph{isolation problem}. This phenomenon is valid for all types of white-box test generation techniques, thus handling code dependencies is an active research topic.
    

In traditional manual unit testing, the isolation problem can be solved using test doubles (mocks, stubs or fakes \cite{xunit}). These test doubles replace the original object (and its methods) to be isolated from the unit under test. Isolation frameworks usually build on one of the following approaches: either they inject a proxy object to the unit under test, or they dynamically rewrite the compiled code of the running program on-the-fly to detour the invocation. The usage of isolation is very common in unit testing, yet its combination with automated test generation is not prevalent; only few papers aim to provide a generic solution (e.g., \cite{pasternak,galler-mock,automock,islam-mock}).

\subsection{Our proposed approach}

In this paper, we present an approach that is able to overcome the isolation problem for white-box test generation by automatically transforming the source code of the unit under test (UUT). These transformations ensure that the invocations inside the UUT will only reach out to a parameterized sandbox, where arbitrary behaviors are created with values obtained from the white-box test generator (e.g., a return value for an isolated method). This generalizes the concept of parameterized mocks \cite{mock-gen,tillmann-put}, while provides an alternative for isolation frameworks. The question that motivated our research is the following:

\begin{quote}
    How well do automated source code based transformations alleviate the isolation problem in white-box test generation?
\end{quote}

We implemented the approach in a ready-to-use tool called AutoIsolator using Microsoft Pex (also known as IntelliTest)~\cite{pexwb} as test generator. We designed a large-scale evaluation involving 10 open-source projects from GitHub. We measured the statement and branch coverage reached by Pex with and without the transformations to decide, whether the approach is capable of alleviating the isolation problem in the given context. Also, to gain better understanding of the practical applicability of our approach, we measured the additional time required by the transformations in the scope of the test generation process.

\subsection{Results}

The results from our evaluation show that automated source code transformations are able to improve white-box test generation by tackling the isolation problem. If AutoIsolator was able to help (i.e., the isolation problem hindered the test generation), then it improved the code coverage with \~50\% on average. Also, the time required by AutoIsolator was on the same order of magnitude as the test generation of Microsoft Pex, thus it does not disproportionately increase the whole time spent with testing. Our results might also indicate that the problem of external dependencies is an important factor, but not the dominant issue for white-box test generators in the open-source wild. Note that we have made the dataset and the analysis script publicly available~\cite{dataset}.


The paper is organized as follows. Section~\ref{sec:motivation} presents a motivating example for the isolation problem in white-box test generation. Section~\ref{sec:background} discusses related research. Section~\ref{sec:approach} shows the automated source code transformations we elaborated to alleviate the isolation problem. Section~\ref{sec:impl} introduces our implementation for Microsoft Pex, while Section~\ref{sec:eval} describes a large-scale evaluation of the approach and the tool. Finally, Section~\ref{sec:conclusions} concludes the paper.
\section{Motivational example}\label{sec:motivation}

In this section, we present a complete practical example, which aims to show the problem of isolation in white-box test generation. First, the method under test is introduced, then we describe the many factors that could hinder the test generation process. Finally, we show how a concrete test generator fails to cover all statements in the code under test. Note that in Sect.~\ref{sec:impl}, we show how automated isolation can improve the coverage achieved via the implemented syntax tree transformations.

Throughout this paper we define the unit under test (UUT) as a single object-oriented class~\footnote{Note that our approach could be used for UUT defined in courser granularity (e.g. classes in the same namespace).}. Also, for each measurement we perform -- including this example -- we measure the code coverage achieved for a single method under test inside the UUT.

If an invocation (or other member access) reaches to a method that is defined outside the UUT, we mark it as external. These external methods (or members) may induce unexpected behavior or unwanted side-effects when they are invoked. Throughout the unit testing phase, these invocations (and member accesses) should be isolated and replaced with something else to avoid such issues. The replacement can return any kind of custom, yet simplified behavior that is defined by the user (e.g., based on a given specification), or extracted automatically from an environment model (e.g., a state machine).

Let us consider \verb|TransferMoney| as the method under test for the example with its source code found in Listing~\ref{lst:running-ex}. This method implements a simplified money transfer workflow consisting of multiple atomic steps. The static method takes three parameters in respective order: the current user token, the amount to transfer, and the destination account. First, the method checks whether the amount to transfer is larger than zero. Then, a database query is run to verify that the user has enough balance to perform the transfer. Finally, an external service called \verb|TransferProcessor| is called to move the money to the other account. If the transfer was successful the method returns \verb|true|, while \verb|false| otherwise.

\begin{figure}[!htb]
    \begin{lstlisting}[caption={An example method that implements a simple money transfer logic.},label=lst:running-ex]
public static bool TransferMoney(Token userToken, long amount, Account destination)
{
    if (amount <= 0) throw new Exception("Invalid amount to transfer");
    int balance = DB.RunQuery<int>("GetBalance", userToken);
    if (balance < amount) throw new Exception("Not enough balance");
    TransferProcessor tp = new TransferProcessor(userToken);
    ProcessedTransfer pt = tp.Process(amount, destination);
    if (pt.IsSuccess) return true;
    return false;
}
    \end{lstlisting}
\end{figure}

This seven lines of code -- without performing any isolation -- has multiple difficulties for white-box test generators. First, unit test generators should not have any access to databases (line 4), because data access layers could be too complex to explore or test generators could delete parts of the database. Second, constructing external objects (\verb|TransferProcessor|) inside the unit under test is an action that should be avoided (line 6), because they could call unknown, external services and thus they could be unintentionally accessed by the test generator. These may also lead to unwanted side effects for the test generation process. Finally, in line 7, there is an invocation to an external method (\verb|Process|), which has a boolean return type. The value returned is tested inside the branching condition of the subsequent line. By default the called method's code will be traversed and its return values (for each possible case) will be calculated based on the traversal's outcome. However, if this method is not fully implemented yet, or it depends on other complex algorithms that are not explorable, the test generator may fail to provide inputs to trigger both \verb|true| and \verb|false| return values.

If there is no isolation provided for the database connection, the white-box test generator will traverse until the code raises an \verb|SqlException| at line 4 indicating that the database connection is not available from the host, from where the invocation is started. Thus, only two lines (lines 3 and 4) of code will be covered by default. To demonstrate this, we implemented the method under test in C\# and executed IntelliTest on it. The generated test inputs and their respective outcomes are shown in Table~\ref{tab:running-ex1-noisolation}.

\begin{table}
    \centering
    \setlength{\tabcolsep}{3pt}
    \caption{The set of test inputs generated by IntelliTest for the example method without isolation.}
    \label{tab:running-ex1-noisolation}    
    \begin{tabular}{c c c c C}\toprule 
        \emph{ID} & \emph{userToken} & \emph{amount}  & \emph{destination} & \emph{result} \\
        \midrule
        T1 & \verb|null| & 0 & \verb|null| & Exception \\        
        T2 & \verb|null| & 1 & \verb|null| & SqlException \\
        \bottomrule
    \end{tabular}
\end{table}

The simplest way of manually isolating a problematic external dependency is to replace the method with a single stub that returns a fixed value. In the example's case, to avoid access to the database, we replaced the \verb|RunQuery| method with a stub that returns 0 for every invocation. With this simple step, we were able to isolate the database access from the test generator, however we also constrained its behavior in terms of test generation. The tests generated by Pex described with their inputs and outcomes are shown in Table~\ref{tab:running-ex1-results}. There are two generated tests achieving only 50\% statement coverage. In order to step through the balance checks the test generator should be able to dynamically change the return value of the \verb|RunQuery| method. However, as described in Sect.~\ref{sec:intro}, creating such stubs or mocks is far from trivial and could be time-consuming -- if possible at all.

\begin{table}
    \centering
    \setlength{\tabcolsep}{3pt}
    \caption{The set of test inputs generated by IntelliTest for the example method with manual stubs.}
    \label{tab:running-ex1-results}    
    \begin{tabular}{c c c c C}\toprule 
        \emph{ID} & \emph{userToken} & \emph{amount}  & \emph{destination} & \emph{result} \\
        \midrule
        T1 & \verb|null| & 0 & \verb|null| & Exception \\        
        T2 & \verb|null| & 1 & \verb|null| & Exception \\
        \bottomrule
    \end{tabular}
\end{table}

To enable the white-box test generator to reach 100\% statement coverage, one should define all possible outcomes (return values and possibly side-effects) for the isolated, external methods. One can achieve this via parameterized mocks, where the test generator can decide what type of behavior to induce in the called methods. Also, in the current example, injection of the two, externally-typed objects (\verb|TransferProcessor| and \verb|ProcessedTransfer|) is not possible due to the restricted parameter list (i.e., they are instantiated inside the method body). Thus, first a testability refactoring is required (which might influence other parts of the code as well), and then the fake methods should be defined with the simplified behavior. This might take significant amount of time and effort, as these should be performed for each method under test. Therefore our goal was to recommend an automatic approach that does not require manual refactoring or mock definitions.
\section{Related work}\label{sec:background}

This section presents various related works from different aspects that include: unit testing in isolation, mock generation, source code transformation-based isolation, and behavior definitions inside the isolated dependencies.

\paragraph{\textbf{Unit isolation for test generation}} There are approaches, which are specific to what they isolate (e.g., \cite{moda,arcuri-ase}), while others use concrete, proxy-based isolation frameworks for this purpose (e.g., \cite{arcuri-mock,mock-gen}). The drawback of using a domain or tool specific approach is that it introduces several constraints to the isolation process:
\begin{itemize}
    \item \emph{Configurability of behaviors:} Configuration of test double behaviors is a crucial part of isolation. White-box test generators require as much configurability as possible. However, each domain/framework has its own limitations on how the behavior of the isolated methods or members can be configured, which may restrict the test generation process itself.
    \item \emph{Applicability:} Using proxy-based isolation frameworks restricts the applicability of isolation in several scenarios (e.g., Mockito does not support static classes, constructors), which may be required for legacy systems, where white-box test generators arise.
    \item \emph{Maintainability:} Domain-specific mock generation requires the extension of the test generator for each domain, while using a given isolation framework demands for tool-specific extensions. Both hinders the maintainability of the test generator.
\end{itemize}

\paragraph{\textbf{Mock generation}} There are several approaches dealing with the automated handling of external dependencies. Tillmann \emph{et al.} proposed the idea of mock generation \cite{mock-gen} along with an evaluating case study \cite{mock-gen-fs} showing promising results. The approach they presented uses explicit interface definitions to generate mocks with basic behavior (not including side effects), while our approach uses extracted data from the invocation sites (instead of interfaces). Galler \emph{et al.} \cite{galler-mock} use contract specifications to generate mock objects concerning the contracts defined manually by the users. It is unclear whether the approach they presented is usable for white-box test generation as well. Arcuri \emph{et al.} \cite{arcuri-mock} extended EvoSuite, a well-known search-based white-box test generation tool to 1) directly access private API methods and 2) create framework-specific mock objects automatically. Pasternak \emph{et al.} \cite{pasternak} proposed a symbolic execution-based test generation engine that produces mocks from previous program executions alongside the generated tests. Our approach does not require previous program executions. Alshahwan \emph{et al.} \cite{automock} presented the \textsc{AutoMock} approach, which performs on-the-fly replacement of external calls into automatically generated mocks. We considered it as an alternative approach to ours, but due to the dependencies on mocking libraries, it could face limitations on large-scale software. On the other hand, their approach of inferring postconditions into mock objects is interesting and can be an extension for our approach.

\paragraph{\textbf{Transformation-based isolation}} Taneja \emph{et al.} \cite{moda} proposed the \textsc{MoDA} approach to generate mock objects for automated test generation for database applications. It generates mocks that mimic the behavior of a database, and the database invocations with queries are automatically replaced on source code level. This approach is somewhat similar to our transformation approach, yet ours intends to provide a more general solution to the problem by handling various type of calls to external dependencies. Klikovits \emph{et al.} \cite{semi-purification} introduced the idea of semi-purification that replaces external functional calls and global variables in C\# functions to eliminate the dependencies. These are replaced with variables that are defined as new parameters for the method. This approach is also similar to ours, however instead of replacing invocations with single variables, our approach creates new invocations to the sandbox that can be extended with arbitrary behavior (including side-effects). 

\paragraph{\textbf{Interactions with dependencies}} Jeon \emph{et al.} \cite{jeon} presented an approach that can synthesize models for external API accesses. Their solution is designed to be used for symbolic execution. Thus the engine can utilize the behavior in the generated framework model. This technique is similar to our sandbox concept, but ours does not require predefined empty declarations, instead, it automatically generates the code. However, the concept of behavior generation into these methods can be employed for our approach as a future extension. Similarly Havrikov \emph{et al.} \cite{zeller-structured} elaborated an approach to support unit test generation with structured interactions between the unit and its dependencies. They integrated an XML generator tool to generate meaningful behavior into the mocks to avoid generating false positive tests. This approach could be integrated with the sandbox concept of our technique.

\medskip Previously, we elaborated a preliminary approach \cite{perpol}, which is able to automatically isolate external dependencies in test generation using heavily invasive source code transformations. However, that approach suffered from several major problems on large programs due to the radical code transformations it requires. Thus, we restarted our research from the basics that led us to the approach presented in this paper.

\section{Approach}\label{sec:approach}

\subsection{Overview and main concepts}

\begin{figure*}[!ht]
    \centering
    \includegraphics[width=0.8\textwidth]{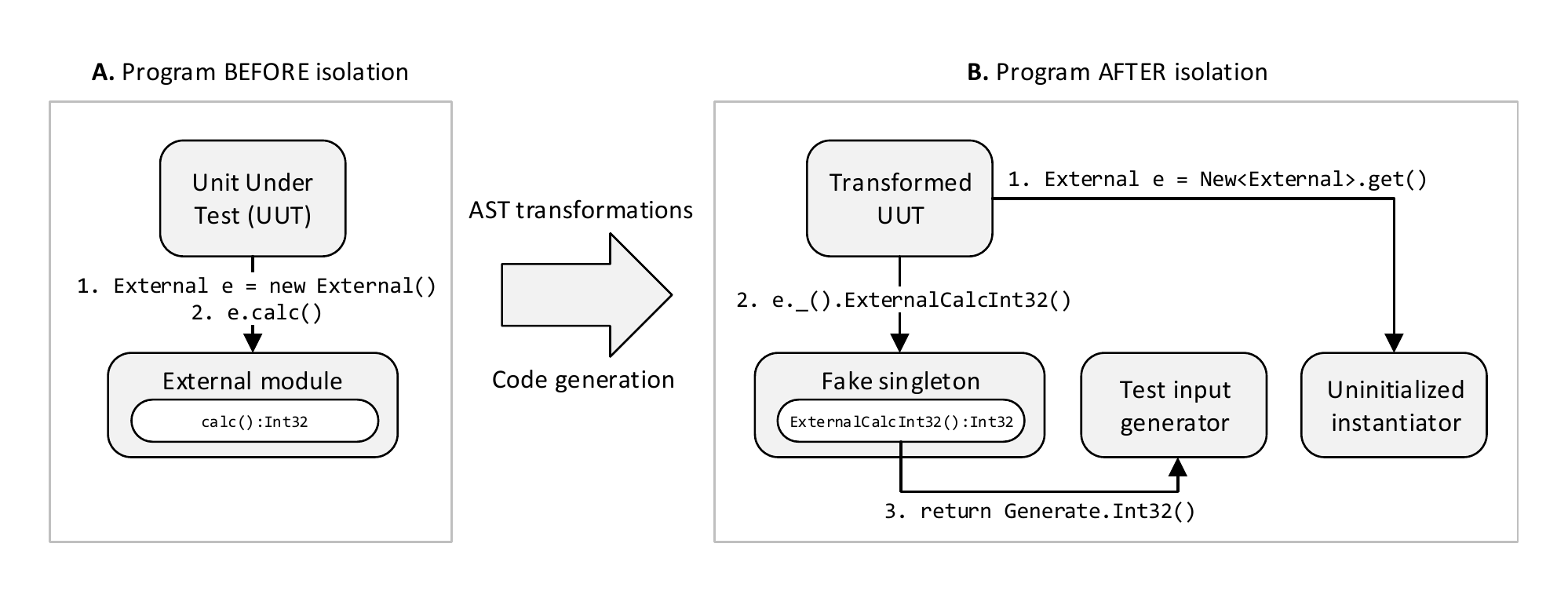}
    \caption{An overview of the automated isolation approach for white-box test generation with example code snippets.}
    \label{fig:methodology-overview}
\end{figure*}

Figure~\ref{fig:methodology-overview} shows the overview of our proposed approach for automated isolation in white-box test generation. The input required by the technique is the fully qualified description of the unit under test (which can contain multiple classes or even modules). The left hand side in the figure visualizes a simple program, which reaches to an external service (\verb|External|) by instantiating it and then calling its method \verb|calc|. The right hand side presents the structure of the unit under test after the automated isolation process that consists of special abstract syntax tree (AST) transformations and code generation. The transformed unit instantiates the external service as an \emph{uninitialized object} and calls into a \emph{Fake singleton object} instead of the original object in the memory. This allows the test input generator to provide return and state changing values in the replaced method. We describe the main concepts of the workflow below in detail.

The transformation algorithm (Algorithm~\ref{alg:overview}) uses the source code and a unit definition (with a sequence of fully qualified name as an input). First, it parses the source code to obtain the syntax trees and the project being worked on, then performs the two main types of syntax tree transformations: member access and object creation. For each successful transformation a fake implementation is generated (retrieving concrete values from the test generator), along with other basic mandatory environment extensions described later in this section. Finally, the generated syntax trees are added to the project, while the original syntax trees are replaced with the transformed ones. If everything has been performed as expected, the project can be compiled again without any errors.

\begin{algorithm}
    \KwIn{Qualified name of the unit under test ($uut$)}
    \KwIn{Source code of the whole project ($sc$)}
    \KwOut{The transformed source code}
    $p \gets parseSourceCode(sc)$\;
    $ti \gets transformMemberAccesses(p, uut)$\;
    $to \gets transformObjectCreations(ti, uut)$\;
    \If{$ti.isSuccess$ \textbf{and} $to.isSuccess$} {
        $f \gets generateFakeCode(ti.Data)$\;
        $e \gets generateBasicEnvironment()$\;
        $p \gets p.replaceOrAddSyntaxTrees(\langle ti, to, e \rangle)$\;
    }
    \Return{$p$}\;
    \caption{AutoIsolator workflow overview}\label{alg:overview}
\end{algorithm}

\subsubsection{Fake singleton}

The Fake object contains all external method or member definitions that are being accessed from the unit under test throughout the isolation process. Each definition represents a single invocation distinguished by a globally unique identifier (thus multiple definitions may exist for the same method but with a unique invocation identifier). The definitions in the Fake singleton contain the custom behavior specified by the test generator (by querying for a given type of values) or even by the user. The advantage of wrapping all instance member definitions into one, globally singleton object is that the injection of this object into the unit under test is much simpler than injecting multiple replacement objects on-demand at each callsite. Note that the isolation of static members is performed using generated static fakes on per class basis as instances are not required to be injected in those cases. See Listing~\ref{lst:example-fake} for a thorough example describing the structure for the generated fake code.

\begin{algorithm}[htb]
	\KwIn{Metadata of member accesses ($md := \{m_0, m_1,\ldots , m_n\}$)}
	\KwOut{Syntax tree of the Fake singleton file ($sf$)}
	$st{\f}, si{\f} \gets \langle\emptyset\rangle$\;
	\For{$i \gets 0$ \textbf{to} $n$} {
		$id \gets extractMemberName(m_i).concat(i)$\;
		$p \gets extractParameterList(m_i)$\;
		$tp \gets extractTypeParameters(m_i)$\;
		$as \gets \langle\emptyset\rangle$\;
		\For{$j \gets 0$ \textbf{to} $p.length$} {
			$t \gets getType(p_j)$\;
			\If{\textbf{not} $isPrimitiveType(t)$} {
				$vg \gets generateValueGenerator(t, tp)$\;
				$as \gets as + generateAssignment(p_j, vg)$\;
			}
		}
		$r \gets generateReturnStatement(m_i)$\;
		$b \gets generateBody(as, r)$\;
		\If{$isStatic(m_i)$ \textbf{or} $isStatic({m_i}.Container)$} {
			$st{\f} \gets st{\f} + generateMember(id, p, tp, b)$\;    
		}
		\Else{
			$si{\f} \gets si{\f} + generateMember(id, p, tp, b)$\;
		}
	}
	$r1 \gets generateStaticFakeClasses(stf)$\;
	$r2 \gets generateSingletonFake(sif)$\;
	\Return{$combineIntoFile(r1,r2)$}\;
	\caption{$generateFakeCode$}\label{alg:fakegen}
\end{algorithm}

Algorithm~\ref{alg:fakegen} describes how the code generation is performed. The algorithm starts from the member access metadata gathered during the transformation process. Then, for each member access ($m_i$), all required information is extracted to generate a fake copy of the member: identifier, and if applicable the parameter list and type parameter list as well. If the member is a method, then all parameters are examined and a value generator statement is emitted and assigned, if the parameter is of a non-primitive type. Last, the same is performed for the return statement (if applicable), and the whole body is assembled into a partial syntax tree. If the member itself, or its container (type) is static, then the member will be emitted to a separate static fake class. Otherwise, the generated member definition is written to the Fake singleton.

\begin{figure}[!htb]
    \begin{lstlisting}[caption={Example describing the structure, how the generated fake code works for instance and static members.},label=lst:example-fake]
public class Example {
    External e = new External();
    int a = e.calc();
    int b = e.calc();
    int c = External.staticCalc();
    return (a+b-c) > 10;
}

public class IsolatedExample {
    External e = New<External>.get();
    int a = e._().ExternalCalcInt32_0();
    int b = e._().ExternalCalcInt32_1();
    int c = FAKE_External.staticCalcInt32();
    return (a+b-c) > 10;
}

public class Fake {
    public int ExternalCalcInt32_0() {
        return Generate.Int32();
    }
    public int ExternalCalcInt32_1() {
        return Generate.Int32();
    }
}

public class FAKE_External {
    public int staticCalcInt32() {
        return Generate.Int32();
    }
}
    \end{lstlisting}
\end{figure}

\begin{figure}
	\begin{lstlisting}[caption={The definition of the isolator method with C\# syntax.},label=lst:isolator-method]
	public static Fake _<T>(this T obj) {
	return Fake.Instance(obj);
	}
	\end{lstlisting}
\end{figure}

\subsubsection{Isolator method}

In order to replace the original member invocations to the fake ones in a static way (without concrete execution), the syntax trees must be modified inside the unit under test. If only the member's name is modified at each invocation, it would cause a compilation error, because the generated fake member (with the given name) is located in an other type (e.g., in the Fake singleton). Thus, the fake container type (for non-static members) must be injected using only syntax tree transformations, which could replace the original type and the generated fake member can be invoked. We achieve the injection of the Fake singleton by using a special \emph{extension method}. There are other approaches that could achieve this (e.g., changing the object name on which these isolated methods are being called), however -- based on our experiences -- they could get too problematic due to their invasiveness in the code.

Extension methods are special programming language constructs that allow developers to extend the behavior of any type without modifying the original, thus these methods can be called on objects having the desired type. Compilers use a dispatcher to enable the syntactic sugar of invoking the extension methods on the original type instance. Most modern languages support extension methods (e.g., Java, C\#, Scala, Kotlin) either out-of-the-box, or with the use of advanced libraries.

Our approach requires a single extension method (the isolator method) denoted with an underscore, which practically injects the Fake singleton object into the unit under test at the callsite. We achieve this by attaching the isolator to all types in the program with the use of the generic signature for an extension method shown in Listing~\ref{lst:isolator-method} (using C\# syntax).

\subsubsection{Uninitialized instantiation}

In order to avoid constructor invocations acting as a leakage from the isolated unit under test, our approach transforms those object creations. We applied the concept of uninitialized objects usually employed for serialization purposes: when deserializing an object into memory, the processing logic must allocate the required size of empty memory space, where the whole object will be stored. That object is called uninitialized, when its memory is allocated, but still empty. The use of unitialized objects is supported in most modern programming languages (e.g., JVM-based languages, C\#). We use the references to these unitialized instances throughout the isolated unit under test to prevent any unwanted logic to execute in any of the constructors. Note that due to the isolator method introduced before, members of these instances are never invoked, thus no error will occur. However, the reference to the allocated memory is passed to the generated code in the Fake singleton (as the isolator extension method knows on which object reference it is used), where it is used for uniquely identifying the object itself. This way we ensure that the generated code can provide various behavior for the same object in different states.

\subsection{Member accesses}

In order to replace the original method invocations in the unit under test, we use isolating abstract syntax tree (AST) transformations. Algorithm~\ref{alg:memberaccess} shows how the transformation is performed for each syntax tree in the unit under test. These are minimally invasive by design, thus they do not change any other behavior in the program. First, the type information is retrieved for both the member itself, and for the callsite as well. Then, each member access node in each syntax tree of the UUT is checked in terms of: whether i) the caller of the member is not a subtype of the member's container, and ii) the member is contained by an external type (outside of the UUT). When both requirements are met, the transformation will take place for the given node. Our proposed approach performs two main changes at each instance invocation (or member access that have no arguments).

\begin{itemize}
    \item \emph{Identifier replacement:} As the Fake singleton contains all of the replaced method definitions (with a unique invocation identifier), we have to ensure that the generated method code has a unique name, while maintaining the essential data for the users. We concatenate the following data in respective order for identification: type name, method name, parameters' types, return type, unique callsite identifier.
    \item \emph{Isolator method insertion:} The call to the isolator method (denoted with an underscore -- as defined in Listing~\ref{lst:isolator-method}) is inserted before the original member identifier node in the AST. This enables the injection of the Fake singleton into the unit under test.
\end{itemize}

\begin{algorithm}[htb]
	\KwIn{Syntax trees of the unit under test ($sts := \{st_0, st_1,\ldots , st_n\}$)}
	\KwOut{Syntax trees after transformation ($trs$)}
	$trs \gets \langle\emptyset\rangle$\;
	\For{$i \gets 0$ \textbf{to} $n$} {
		$ms \gets extractMemberAccessNodes({st_i})$\;
		\For{$j \gets$ 0 \textbf{to} $ms.length$} {
			$s \gets retrieveTypeIn{\f}ormation(ms_j)$\;
			\If{\textbf{not} $baseTypeO{\f}({ms_j}.caller, {ms_j}.container)$ \textbf{and} $isExternal(ms_j)$} {
				$nm \gets extractMemberName(s).concat(j)$\;
				\If{$isStatic(ms_j)$ \textbf{or} $isStatic({ms_j}.Container)$} {
					$ma \gets generateStaticMemberAccess(s,nm,ta)$\;
					$st_i \gets st_i.replaceNode(ms_j, ma)$\;
				}
				\Else{
					$ma \gets generateMemberAccess(s,nm,ta)$\;
					$st_i \gets st_i.replaceNode(ms_j, ma)$\;
				}
			}
		}
		$trs \gets trs + st_i$\;
	}
	\Return{trs}\;
	\caption{$transformMemberAccesses$}\label{alg:memberaccess}
\end{algorithm}

In terms of static methods or members, we only perform two simple identifier transformations: i) the static type's name receives a prefix to identify the replacement, and ii) the method or member identifier is transformed the same way as described for instance members.

\begin{figure*}[t]%
	\centering 
	\subfloat[An invocation AST before isolating transformations.]{
		\includegraphics[width=0.45\textwidth]{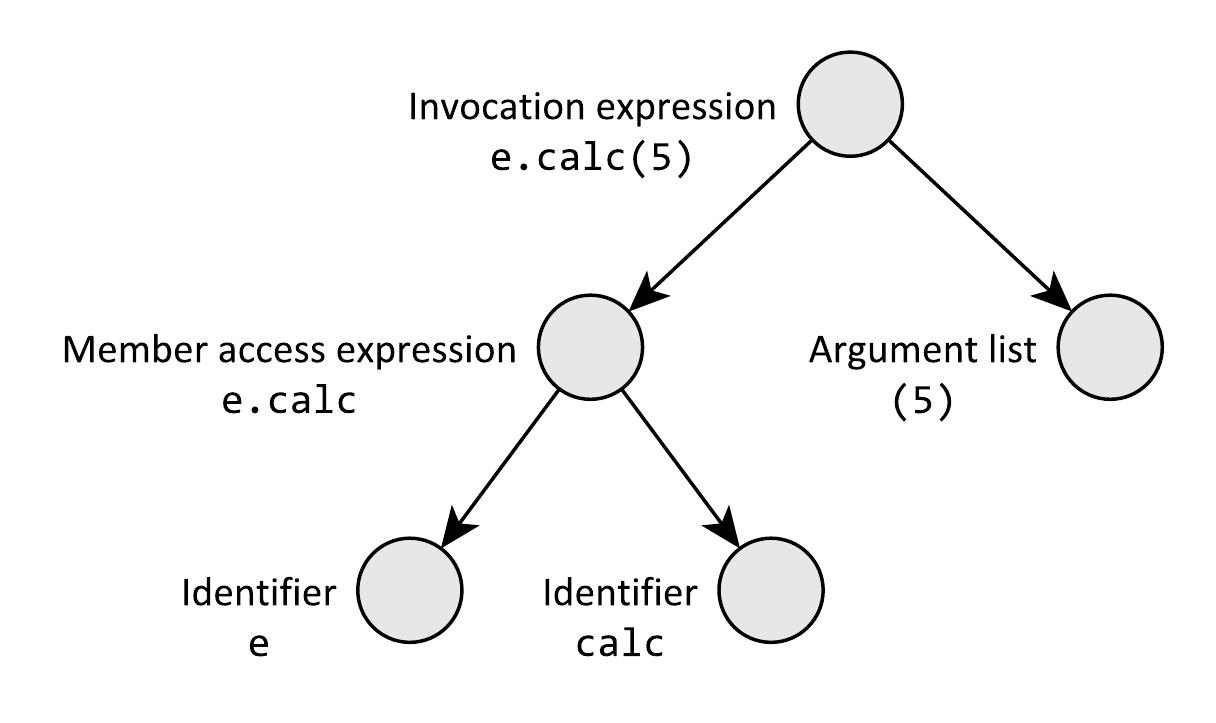}
		\label{fig:ast-memberaccess-original}
	}\qquad 
	\subfloat[The transformed AST of a method invocation]{
		\includegraphics[width=0.35\textwidth]{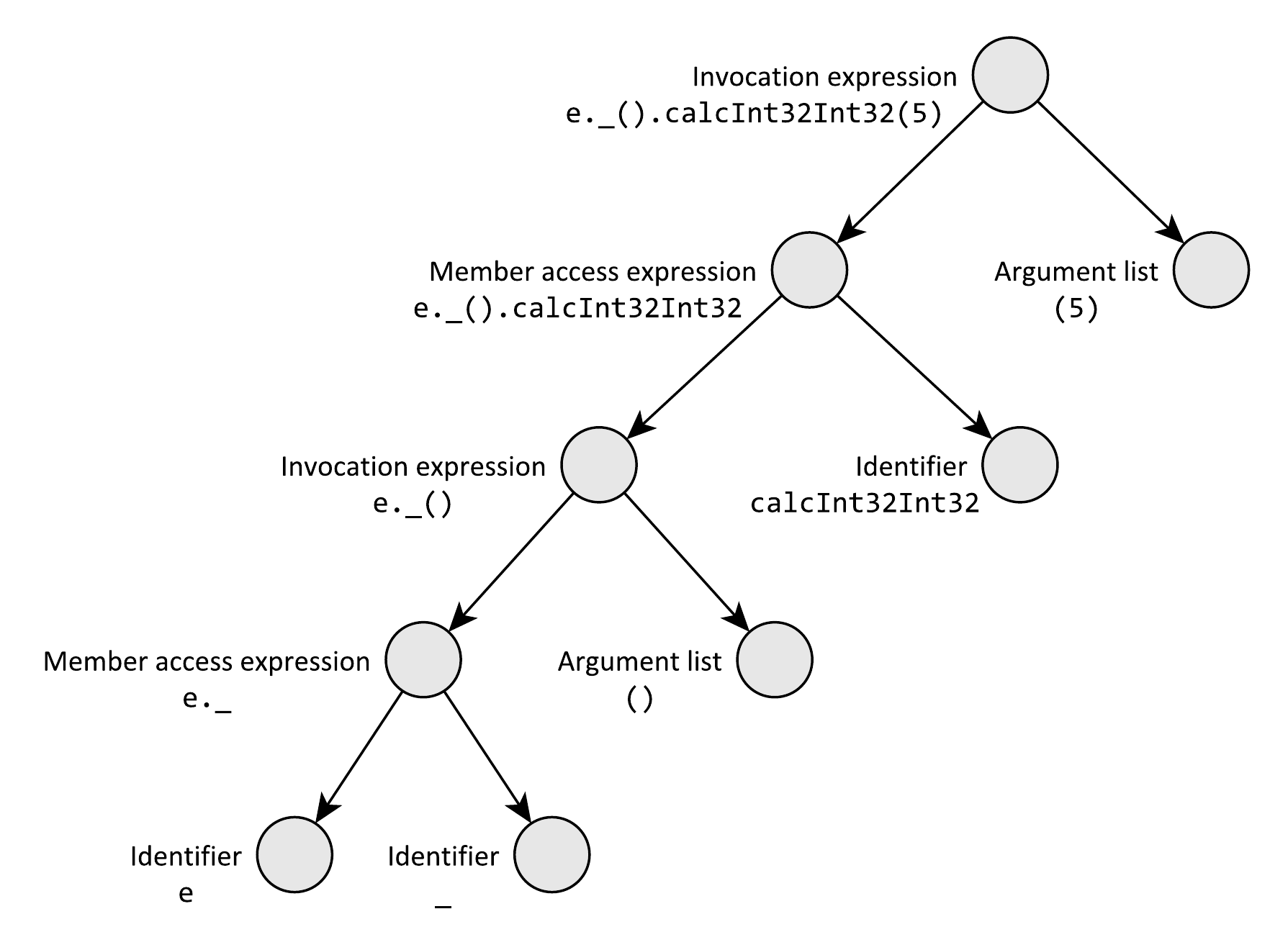}
		\label{fig:ast-memberaccess-after}
	}
	\caption{Example transformation of method invocation} 
	\label{fig:ast-memberaccess} 
\end{figure*}

Figure~\ref{fig:ast-memberaccess-original} shows an AST for a method invocation with a single argument, while Figure~\ref{fig:ast-memberaccess-after} shows the transformed AST after the two modifications.

\subsection{Object creations}

\begin{figure*}%
	\centering 
	\subfloat[An AST of an object creation with arguments.]{
		\includegraphics[width=0.45\textwidth]{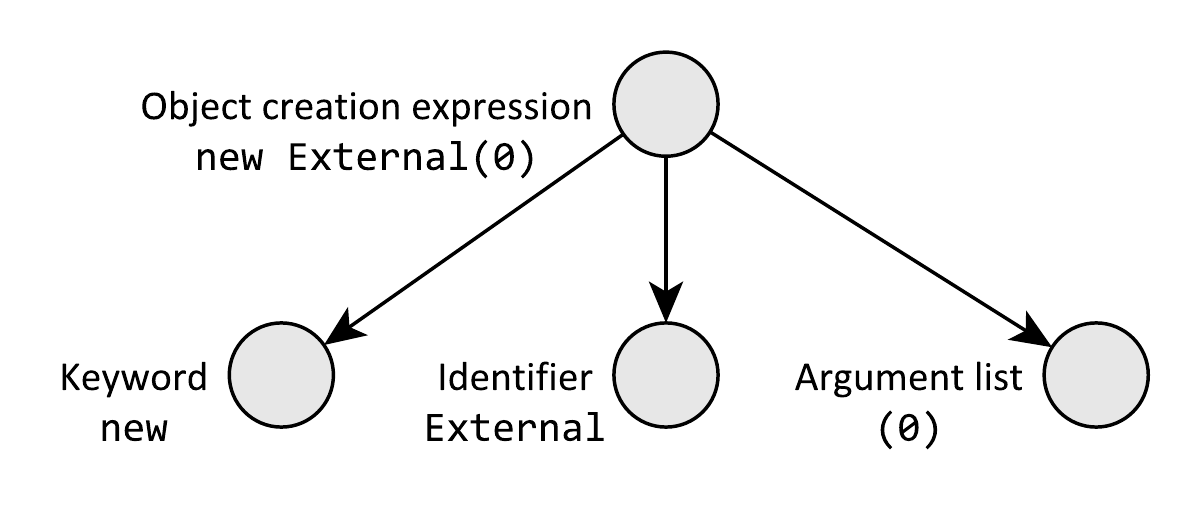}
		\label{fig:ast-objectcreation-original}
	}\qquad 
	\subfloat[Transformed AST of an object creation.]{
		            \includegraphics[width=0.35\textwidth]{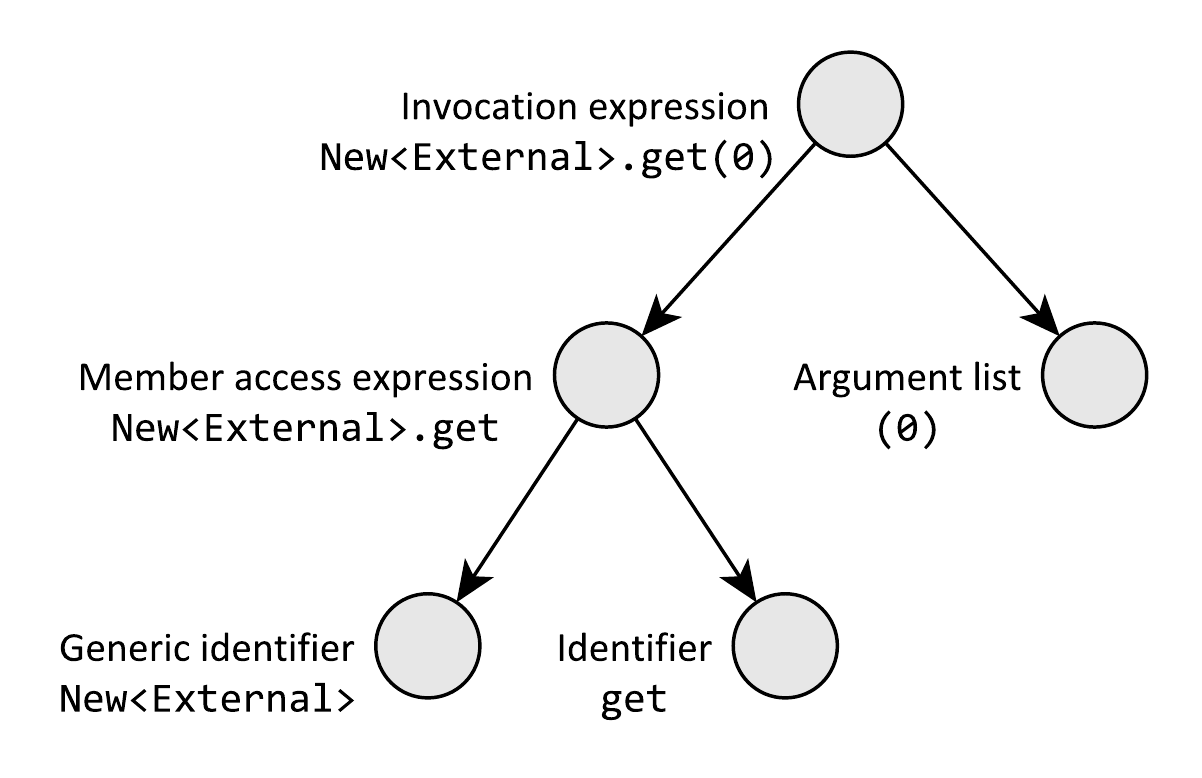}
		\label{fig:ast-objectcreation-after}
	}
	\caption{Example transformation of object creation} 
	\label{fig:ast-objectcreation} 
\end{figure*}

Uninitialized objects are key parts in our approach. This concept ensures that no constructors are invoked throughout the isolated unit under test. In order to achieve this, we apply an AST transformation for each object creation expression. Consider an example AST in Figure~\ref{fig:ast-objectcreation-original}, which shows a simple instantiation of a type with a single argument. The uninitialized object creation is performed in an external utility class, we are required to transform the original object creations into invocations reaching out to this generic utility class (called \verb|New<T>|) and its method (\verb|get|). In Figure~\ref{fig:ast-objectcreation-after}, we present how the constructor invocation found in line 2 of Listing~\ref{lst:example-fake} will be transformed automatically into an invocation to the mentioned utility. Note that all of the constructor arguments are passed to the instantiator method, so that they can be used at later invocations, or for behavior simulation purposes.
\section{Implementation for Microsoft Pex}\label{sec:impl}

We have chosen Microsoft Pex~\cite{pexwb} (IntelliTest) as the test generator for which we implement our approach, because it is one of the most advanced white-box test generators. Also, the language on which Pex works (C\#) supports all of the main concepts used in the approach out-of-the-box (e.g., no external libraries are required for extension methods). Microsoft Roslyn~\cite{roslyn} provides an easy-to-use API for syntax tree transformations and in-memory compilations in C\#, thus we decided to use this library for our implementation purposes.

\subsection{Overview}

\begin{figure*}[!ht]
    \centering
    \includegraphics[width=0.75\textwidth]{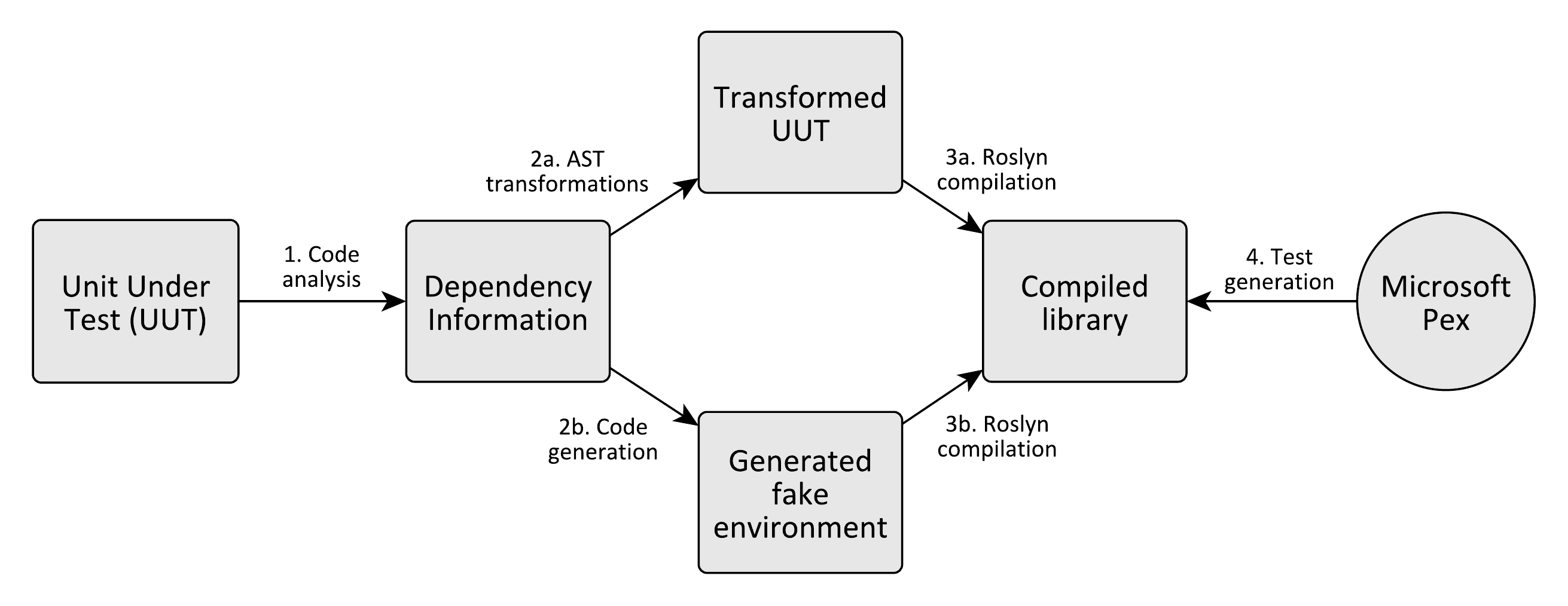}
    \caption{An overview of the workflow in the implemented tool for Microsoft Pex.}
    \label{fig:implementation-overview}
\end{figure*}

To get a quick overview of how our implementation -- called AutoIsolator -- is designed, we present its workflow in Figure~\ref{fig:implementation-overview}. Our tool retrieves the source code of the unit under test by opening the project into memory via Roslyn's project and workspace handling feature. After that -- using code analysis -- the tool extracts the information about all dependencies of the unit under test. Then, based on this information, the UUT is transformed on abstract syntax tree level via Roslyn. In parallel, the required fake method definitions are generated automatically, which are invoked from the transformed UUT. Finally, with Roslyn's in-memory compilation, an isolated dynamically linked library (DLL) is assembled including the changes performed previously. The library also contains generated Parameterized Unit Tests (PUTs~\cite{tillmann-put}) for each externally reachable method in the UUT: they act as starting points for Microsoft Pex's test generation.

\subsection{Current limitations}

\paragraph{Object state handling} Currently the tool does not support tracking the states of objects that have types defined outside of the unit under test. By disregarding the object state the values returned from their isolated methods are quasi-independent, and thus the test generator can choose arbitrary values.

\paragraph{Operators on external objects} Operators are not treated in the isolating transformations at all, which causes a runtime issue when the concrete execution applies the operator (e.g., equality) on an uninitialized object. This is an important future fix to consider.

\paragraph{Special classes as a unit under test}  The current version of the tool does not support generic or nested classes as the unit under test. Generic classes are not supported by Microsoft Pex either, thus this is an external limitation. 

\paragraph{Extension methods} The tool does not support extension methods as a starting point of the test generation, because they can be applied on other types than they are defined in, which makes the definition of the unit under test unclear.

\subsection{Isolation example}

We have presented a motivating example in Sect.~\ref{sec:motivation} that has shown the issues with external dependencies during white-box test generation. To demonstrate the capabilities of our previously presented automated isolation approach, we executed AutoIsolator to enhance the test generation process of Pex for that single unit and method under test.

\begin{figure*}[!htb]
    \begin{lstlisting}[caption={The automatically isolated money transfer method from Section \ref{sec:motivation} and its related generated assets.},label=lst:running-ex-impl]
public static bool TransferMoney(Token userToken, long amount, Account destination)
{
    if (amount <= 0) throw new Exception("Invalid amount to transfer");
    int balance = FAKE_DB.RunQuery__0_<int>("GetBalance", userToken);
    if (balance < amount) throw new Exception("Not enough balance");
    TransferProcessor tp = New<TransferProcessor>.Instance();
    ProcessedTransfer pt = tp._().Process__1_(amount, destination);
    if (pt._().MemberIsSuccess__2_) return true;
    return false;
}
internal static partial class FAKE_DB
{
    internal static int RunQuery__0_<T>(string queryName, params object[] args)
    {
        return PexChoose.Value<int>("a");
    }
}
internal partial class Fake
{
    internal ProcessedTransfer Process__1_(long amount, Account destination)
    {
        return New<ProcessedTransfer>.Instance();
    }
    public bool MemberIsSuccess__2_
    {
        get
        {
            return PexChoose.Value<bool>("b");
        }
    }
}
    \end{lstlisting}
\end{figure*}

The transformed unit under test and the related generated code can be found in Listing~\ref{lst:running-ex-impl}. Note that the generated code differs compared to Listing~\ref{lst:example-fake} as the concrete implementation uses Pex-specific instructions. In line 4 the transformed invocation to the database retrieves an arbitrary integer value from the Pex engine (instead of having a stub with fixed value). In line 6, instead of creating a normal instance of \verb|TransferProcess|, an uninitialized memory space is allocated to that reference: this avoids the possible unwanted side-effects of calling external constructors. Then, line 7 creates an uninitialized instance of \verb|ProcessedTransfer| to avoid calling an external method. Finally the result's boolean property is checked in the branching condition. The property's value is again provided by the Pex engine and thus both outcomes can be covered. We have executed Pex on the transformed unit under test, which resulted in the generated tests and their respective outputs found in Table~\ref{tab:running-ex2-results}. Note that two new tests were generated compared to other cases by which the statement coverage increased from 50\% to 100\%. The new tests were using the arbitrary values that were generated by Pex (see columns \verb|a| and \verb|b| that are representing the generated values from lines 15 and 28, respectively).

\begin{table*}
    \centering
    \setlength{\tabcolsep}{3pt}
    \caption{The set of test inputs generated by Pex for the isolated example method.}
    \label{tab:running-ex2-results}    
    \begin{tabular}{c c c c c c C}\toprule 
        \emph{ID} & \emph{userToken} & \emph{amount}  & \emph{destination} & \emph{a} & \emph{b} & \emph{result} \\
        \midrule
        T1 & \verb|null| & 0 & \verb|null| & - & - & Exception \\        
        T2 & \verb|null| & 1 & \verb|null| & - & - & Exception \\
        \textbf{T3} & \verb|null| & 861 & \verb|null| & 862 & \verb|true| & \verb|false| \\
        \textbf{T4} & \verb|null| & 861 & \verb|null| & 862 & - & \verb|true| \\
        \bottomrule
    \end{tabular}
\end{table*}
\section{Evaluation}\label{sec:eval}

Empirical evaluation of a novel technique and tool is crucial to demonstrate its benefits. In this section, we present the design and the results of our evaluation of AutoIsolator on randomly selected open-source projects.

\subsection{Goal and method}

The goal of this experiment is twofold. First, we would like to demonstrate that the approach implemented in AutoIsolator can improve white-box test generation by automatically isolating external dependencies. Second, we would also like to show that the implemented approach does not require significant amount of additional time during test generation, thus it would be convenient to apply in practice. Based on these two requirements, we formulate the following research questions for our experimental evaluation.

\paragraph{RQ1} How the implemented automated isolation mechanism improves the statement and branch coverage reached by generated white-box tests?
\paragraph{RQ2} On what extent does the implemented approach increase the time spent with the whole test generation process?

\subsection{Experiment planning}

\subsubsection{Variable selection}

In terms of collected data, we separate independent and dependent variables \cite{se-exp} (detailed in Table~\ref{tab:variables}). There were four main independent variables along with 12 other dependent ones. The first main independent variable is the method (Method) for which the white-box tests are generated. The second main independent variable takes a value whether there is isolation performed by our tool or simply Pex was executed alone on the method (IsIsolated). This is the chosen factor for our empirical evaluation with the mentioned two treatments. We also define two descriptive independent variables for the details of the object being executed (CC, LoC).

\begin{center}
    \begin{table*}[htb]
    \caption{Independent and dependent variables defined for the experiment.}\label{tab:variables}
    \centering
    \begin{tabular*}{1\textwidth}{l l p{11.5cm} l}
    \toprule
    \textbf{Cat.} & \textbf{Name} & \textbf{Description} & \textbf{Type} \\
    \midrule
     \multirow{4}{*}{\rotatebox{90}{\textbf{Independent}}}
            & IsIsolated & Represents how the test generation is performed. We define two treatments: executing Pex without or with AutoIsolator. & Factor \\
            & Method & The method to execute including its containing class and project. & Factor \\
            & CC & The cyclomatic complexity of the method. & Numeric \\
            & LoC & The non-commenting source lines of code found in the method. & Numeric \\
    \midrule
     \multirow{12}{*}{\rotatebox{90}{\textbf{Dependent}}}
            & SC & Statement coverage of the method under test. & Numeric \\
            & BC & Branch coverage of the method under test. & Numeric \\
            & TC & The total number of generated tests. & Numeric \\
            & PWC & The number of warnings found in the generated Pex report by type. & Combined \\
            & PEC & The number of errors found in the generated Pex report by type. & Combined \\
            & PB & The possible boundaries reached by Pex at the end of the test generation. & Factor \\
            & IMethods & The number of external method accesses, i.e. the number of method accesses to isolate. This is a unit-level (class) metric. & Numeric \\
            & IMembers & The number of external property or field accesses, i.e. the number of property or field accesses to isolate. This is a unit-level (class) metric. & Numeric \\
            & TTransformation & The time required by AutoIsolator for transformations & Numeric [s] \\
            & TCodeGeneration & The time required by AutoIsolator for emitting the fake methods, properties and fields. & Numeric [s] \\
            & TCompilation & The time required by AutoIsolator for the compilation of the transformed project. & Numeric [s] \\
            & TTestGeneration & The time required by Pex to generate tests. & Numeric [s] \\
    \bottomrule
    \end{tabular*}
    \end{table*}
\end{center}

In terms of dependent variables, we chose 12 that are observed as the outcomes of our evaluation. First to form an answer for RQ1, we measure statement (SC) and branch coverage (BC) values to decide on the enhancements introduced by AutoIsolator. We have considered other alternatives, such as mutation score, to measure the quality of generated test suites. However, we defined the enhancement as an increase in the explored behavior of the unit under test, thus the bug-finding capability of the generated tests (mostly measured by mutation score) is not in the scope of this evaluation. Also, we measured the number of generated tests by Pex (TC). For later investigation we observed the number of warnings (PWC), errors (PEC) and boundaries (PB -- e.g., maximum paths, timeout) in Pex executions. We also measured how many external method (IMethods) and member accesses (IMembers) are there in each unit that were automatically transformed by AutoIsolator. Note that these two variables are dependent, because they are extracted from AutoIsolator, when executed on a given method. Finally, to answer RQ2 we have measured 4 time-related variables (TTransformation, TCodeGeneration, TCompilation, and TTestGeneration) that are concerned with the time required by AutoIsolator compared to plain Pex-only test generations.

\subsubsection{Object selection}

To improve the external validity of our experiment, we selected projects from an external source. As GitHub is the largest location, where open-source C\# projects are available, we selected 10 repositories from there for our experimental purposes. The selection of these repositories was random from the ones fulfilling the following four criteria.

\begin{itemize}
    \item Has at least 1000 stars on GitHub, which represents the popularity of the project.
    \item Has no relation to user interfaces, as they are not well-suitable for white-box test generation.
    \item Has no relation to mobile applications, as they require special project setups that Roslyn does not support (e.g., .NET Core).
    \item Has no relation to graphics, as they require special dependencies that are not handled by Roslyn.
\end{itemize}

The C\# repositories found on GitHub usually contain multiple projects. To simplify the selection process (e.g., do not drop a repository, which has one UI-related project besides others), we only examined and used the root (main) project from each of these repositories. 

After we selected the 10 projects, we implemented a tool that can extract all methods automatically from a C\# project. However, we made sure that all methods used in the experiment are supported by Pex, thus we excluded the methods that had at least one of the following traits. The filtering process was also fully automated.

\begin{itemize}
    \item The method's container is a nested class.
    \item The method's container is an abstract class.
    \item The method's container is a generic class.
    \item The method is abstract.
    \item The method is an extension method.
    \item The method's accessibility (visibility) is not public.
\end{itemize}

Our selection and filtering procedure yielded the projects and methods found in Table~\ref{tab:selected-projects}. The table also describes how many methods were explored (\textbf{EM}), and how many were applicable for our purposes (\textbf{AM}).

\begin{table*}[t]
    \caption{The randomly selected projects and their methods from GitHub. EM denotes the number of total methods explored, while AM shows how many were applicable for us. The rest of the statistics are regarding with the applicable methods.}\label{tab:selected-projects}
    \centering
    \begin{tabular*}{0.86\textwidth}{l l c c c c c}
    \toprule
    \textbf{Repository} & \textbf{Project} & \textbf{\#EM} & \textbf{\#AM} & \textbf{Mean LoC} & \textbf{Total LoC} & \textbf{Mean CC} \\
    \midrule
    Abot & Abot & 172 & 59 & 11.95 & 705 & 2.03 \\
    Akka & Akka & 2049 & 949 & 13.81 & 13074 & 1.48 \\
    GraphEngine & Trinity.Core & 902 & 386 & 16.86 & 6507 & 2.05 \\
    Humanizer & Humanizer & 444 & 114 & 21.00 & 2394 & 4.21 \\
    ImageProcessor & ImageProcessor & 367 & 194 & 35.99 & 6982 & 2.84 \\
    LiteDB & LiteDB & 600 & 326 & 16.08 & 5307 & 2.07 \\
    NodaTime & NodaTime & 681 & 228 & 15.21 & 3468 & 1.76 \\
    Polly & Polly & 746 & 47 & 9.72 & 457 & 1.13 \\
    Simple.Data & Simple.Data & 719 & 390 & 8.56 & 3338 & 1.55 \\
    TopShelf & TopShelf & 507 & 242 & 10.70 & 2590 & 1.50 \\
    \midrule
    Total & 10 & 7187 & 2935 & - & 44822 & - \\
    \bottomrule
    \end{tabular*}
\end{table*}

We executed AutoIsolator and Pex on each of the methods and found that \emph{2596 methods were successfully executed by both tools out of the total 2935 applicable methods}:

\begin{itemize}
    \item \emph{AutoIsolator error:} In case of 300 methods AutoIsolator yielded an uncompilable code or an uncaught exception was thrown during the execution.
    \item \emph{Pex error}: Pex could not emit compilable source code for the generated tests in case of 39 methods.
\end{itemize}

In the forthcoming sections, we only consider the 2596 successfully executed methods, as they are the comparable ones in terms of coverage and required time. These 2596 methods sum up 38618 lines of code, with a mean of 14.88. The cyclomatic complexity (CC) was 1.84 on average.

\subsubsection{Process}

We designed the procedure of the evaluation to answer both RQs at once. To simplify the definition of the unit under test throughout the study, we decided to use single classes as the unit under tests. This way the same transformations are performed for all methods found in the same class. To measure the coverage and time difference for white-box test generation with and without the automated isolation, we executed Pex on them two times (as described previously in the IsIsolated independent variable): first without any isolation, and then with fully automated isolation. For both types of execution, we measured the same, required variables. To enable a valid data collection procedure, and to avoid unexpected outlier values, we performed each measurement 3 times. We analyzed all the numerical variables and found that the standard deviation among the five sessions were negligible. Thus, \emph{all of the values we analyzed are medians} obtained using a preliminary data merge from the 3 measurement sessions.

\subsubsection{Environment}

We used a cloud virtual machine to perform the automated evaluation, because executing test generation with and without AutoIsolator requires a special environment to run continuously for 33 hours. Also, we had to ensure that running Pex without AutoIsolator is not causing unexpected effects on the file system or any other service on the machine, hence we executed the evaluation with a separate user having a strongly restricted access to the file system and other services. The virtual machine we used was running Windows 10 and had 3.5 GBs of memory along with a dedicated CPU running at 2.4 GHz.

\subsubsection{Data analysis}

We used R 3.4.3 \cite{r} to analyze the outcomes. First, we performed exploratory data analysis to gain insights into the data using basic plots and standard descriptive statistics. Then, we sought the answers to our predefined research questions. We support our answers using tables, charts, and basic statistical methods. These assets were mostly automatically generated from the raw data using the R script we created.

\subsection{Threats to validity}

\subsubsection{Construct}

\emph{Mono-method bias.} We measured improvements introduced by AutoIsolator using coverage metrics only. However, there are critiques about assessing test suites using coverage metrics (e.g., \cite{cov-corr}). Such studies usually state that it is unsafe to use coverage values, as they are not well correlated with the effectiveness of test suites. Although we also considered this threat, our tool and its evaluation focus on improving and alleviating the test generation process, not on the quality of generated test suites. The improvement of a test generation process can be measured by comparing the ratio of the explored and covered code. Nevertheless, using mutation score to assess the quality could be an obvious choice when evaluating different kinds of automated behavior generation algorithms that provide actions inside the sandbox.

\subsubsection{Internal}

\emph{Randomness in test generation.} Although Pex uses dynamic symbolic execution, sometimes its underlying constraint solver uses randomized algorithms, which might affect the outcome of the generated tests. This yields that occasionally the generated test data is not reproducible immediately. We treated this threat by executing every measurement three times and obtained the median value from them. 

\emph{Parameters of test generation.} We used the default settings for Pex and provided no extra parameters. However, by default, Pex uses a meta-strategy to select the best search strategy for exploring the code in dynamic symbolic execution. We did not have control over the meta-strategy algorithm. In this sense, it could happen that the improvements presented previously are produced by a better-chosen strategy caused by the transformations. 

\subsubsection{External}

\emph{Selected projects.} To improve the generalizability of our evaluation results, we selected the objects randomly from the popular C\# repositories on GitHub. However, there is a threat that these projects are not well representing the characteristics of all open-source C\# programs. To further eliminate this threat, replication studies are needed with different projects from different sources.

\emph{Measurements in cloud.} We performed our measurements on virtual machines in a cloud environment. As there can be interferences between the machines on the same host, our results in terms of execution times may be distorted by this effect. We tackled this threat by executing the measurements three times and choosing the median values from them. 

\subsection{Results}

\subsubsection{General overview}

First, we provide a general overview of how Pex was able to perform on the selected open-source methods in terms of the number of generated tests and achieved statement and branch coverage. These results are not only serving as a baseline for the comparison with automated isolation, but they are unique results for Pex itself as well: we are not aware of any public evaluation -- at this scale --  of the performance of Microsoft Pex.

\begin{figure}[!ht]
    \centering
    \includegraphics[trim={0 0.8cm 0 0.8cm},width=\columnwidth]{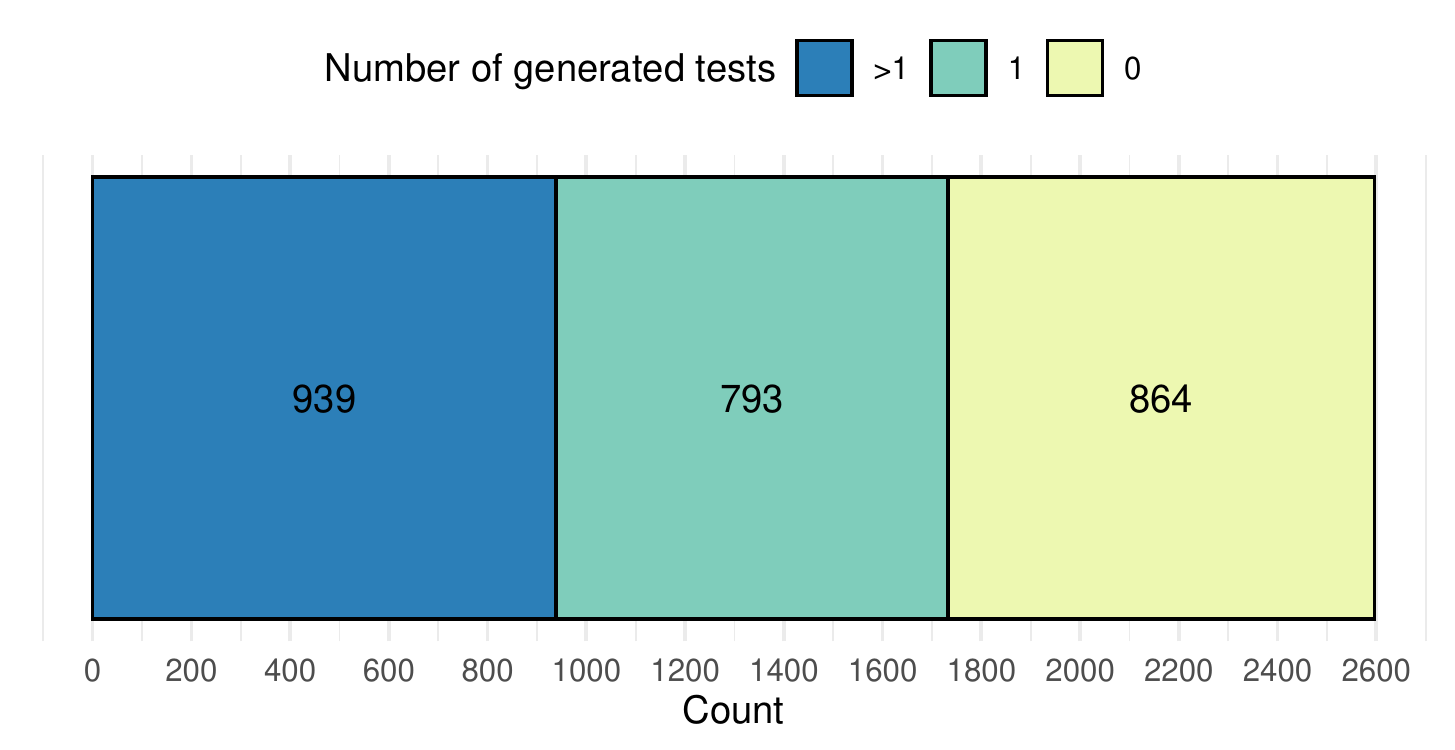}
    \caption{The number of generated tests for the successfully executed 2596 methods (without any isolation), separated into 3 main categories.}
    \label{fig:full-join-testcount}
\end{figure}

Figure~\ref{fig:full-join-testcount} shows how many tests were generated by Pex on the successfully executed methods: in two thirds of the cases there was at least one test case that has been generated. Also, in 36\% of the cases there were more than one generated test cases. For the other third of the cases (864) no tests were generated at all due to various issues around the execution of Pex. We investigated some randomly selected cases among those using Pex's log and found that the reason why Pex could not generate any tests is because it could not instantiate the desired objects through the usable constructors or default factory methods (commonly referred as the object creation problem).

\begin{table}
    \centering
    \setlength{\tabcolsep}{3pt}
    \caption{The basic descriptive statistics of the variables observed during the 2596 executions performed by Pex without any isolation.}
    \label{tab:pexonly-basic-stats}    
    \begin{tabular}{l c c c c c}\toprule 
        \emph{Variable} & \emph{Min} & \emph{Med}  & \emph{Mean} & \emph{Max} & \emph{SD} \\
        \midrule
        TC & 0 & 1 & 2.88 & 60 & 5.55 \\
        SC [\%] & 0 & 66.67 & 52.83 & 100 & 45.18 \\
        BC [\%] & 0 & 92.96 & 56.86 & 100 & 46.64 \\
        \bottomrule
    \end{tabular}
\end{table}

Table~\ref{tab:pexonly-basic-stats} shows the descriptive statistics of the variables about the performance of Pex without any isolation. Based on these, in more than half of the cases Pex generated only a single test case, yet there is a large deviation between methods. In terms of statement coverage (SC), Pex reached a median coverage of 66.67\% with a mean of 52.83\%. Although, the maximum statement coverage reached was 100\%, there was a standard deviation of 45.18\%. The situation is somewhat better for branch coverage (BC): median of 92.96\%, maximum of 100\%, a mean of 56.86\% with a high standard deviation of 46.64\%.

\begin{figure}[!ht]
    \noindent\begin{minipage}{.47\columnwidth}
        \centering
        \includegraphics[trim={0 0.8cm 0 0.8cm},width=\columnwidth]{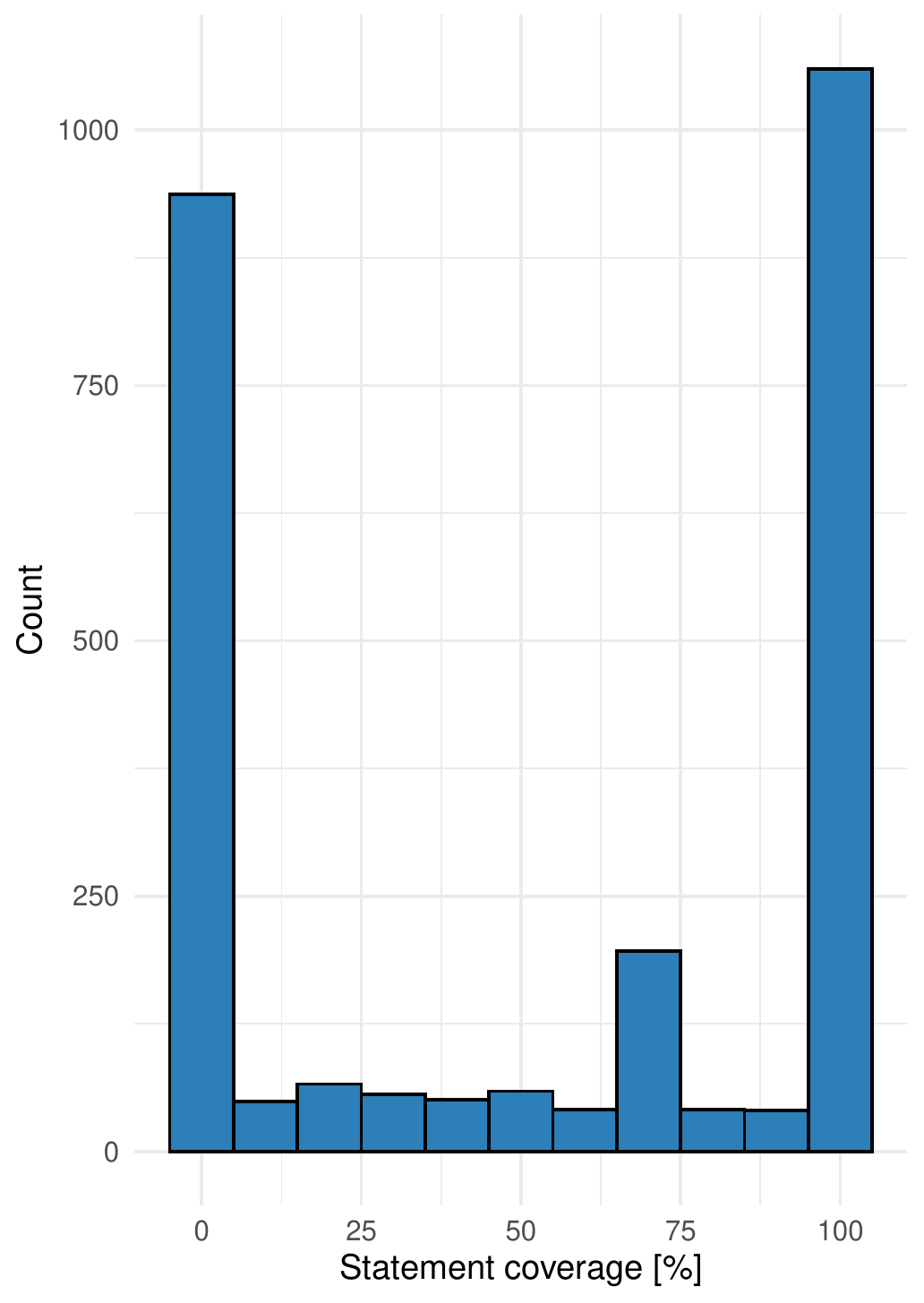}
        \caption{The distribution of statement coverage values achieved by Pex by itself.}
        \label{fig:pexonly-sc-hist}
    \end{minipage}\hfill
    \begin{minipage}{.47\columnwidth}
        \centering
        \includegraphics[trim={0 0.8cm 0 0.8cm},width=\columnwidth]{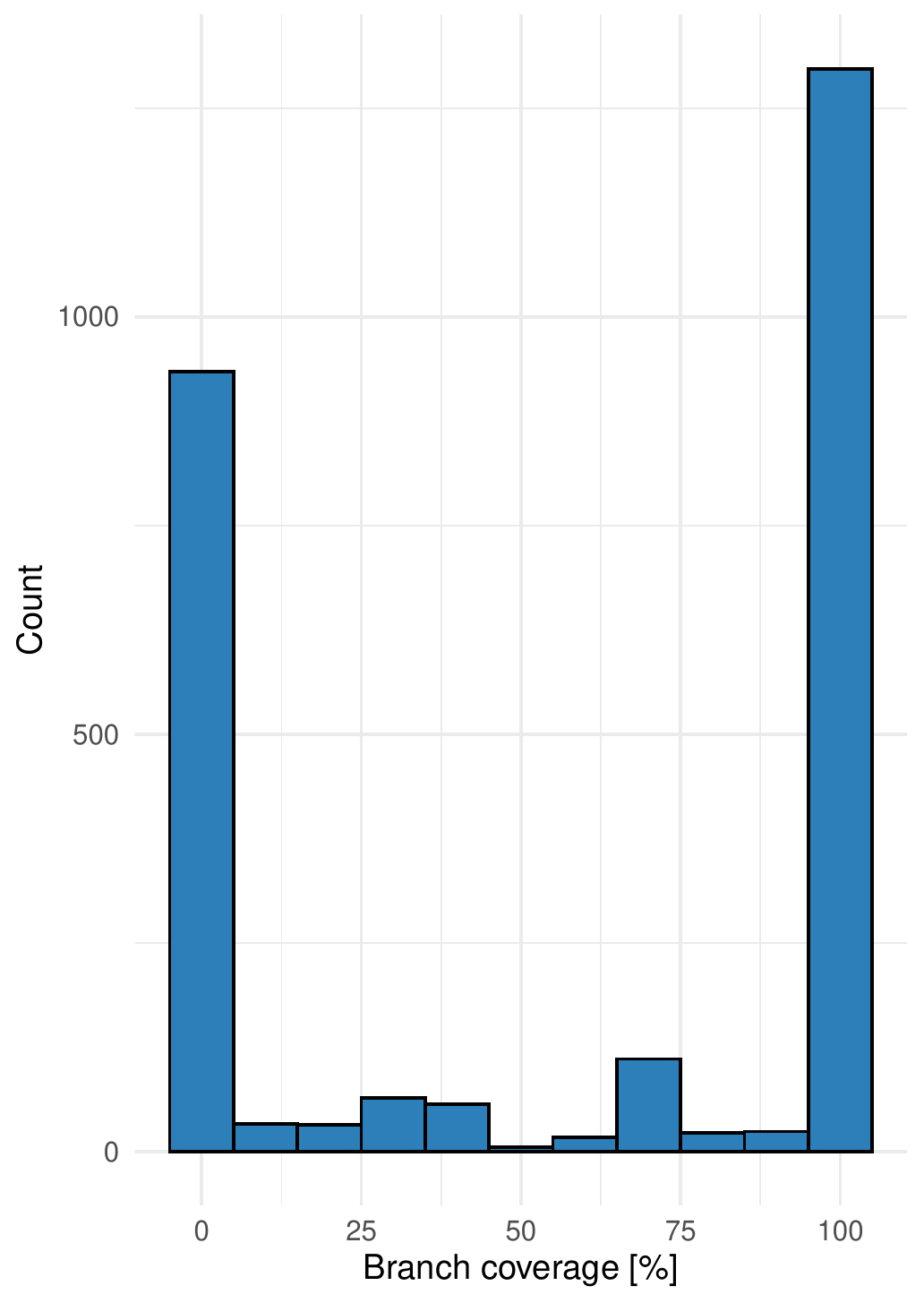}
        \caption{The distribution of branch coverage values achieved by Pex by itself.}
        \label{fig:pexonly-bc-hist}
    \end{minipage}
\end{figure}

A deeper look in the distributions of the statement and branch coverage values (shown in Figure~\ref{fig:pexonly-sc-hist} and Figure~\ref{fig:pexonly-bc-hist}, respectively) yields that in the largest portion of the cases if Pex was able to generate any tests, then it reached 100\% for both coverage types. In terms of statement coverage, there were only 604 cases, where Pex reached more than 0\%, yet less than 100\%.

These results indicate that in most of the cases Pex could generate a single test, which led to full code coverage (i.e., found the golden path). In the second largest portion of the cases Pex reached 0\% code coverage. After analyzing these cases (the corresponding warnings given by Pex), we have found that the main issue -- in correspondence with the number of generated tests -- was that Pex could not instantiate an object, which was needed as a test input (PUT parameter). Although, this is a well-known limitation of white-box test generators, yet it is surprising how this hinders the coverage reached.

\begin{figure}[!ht]
    \centering
    \includegraphics[trim={0 0.8cm 0 0.8cm},width=\columnwidth]{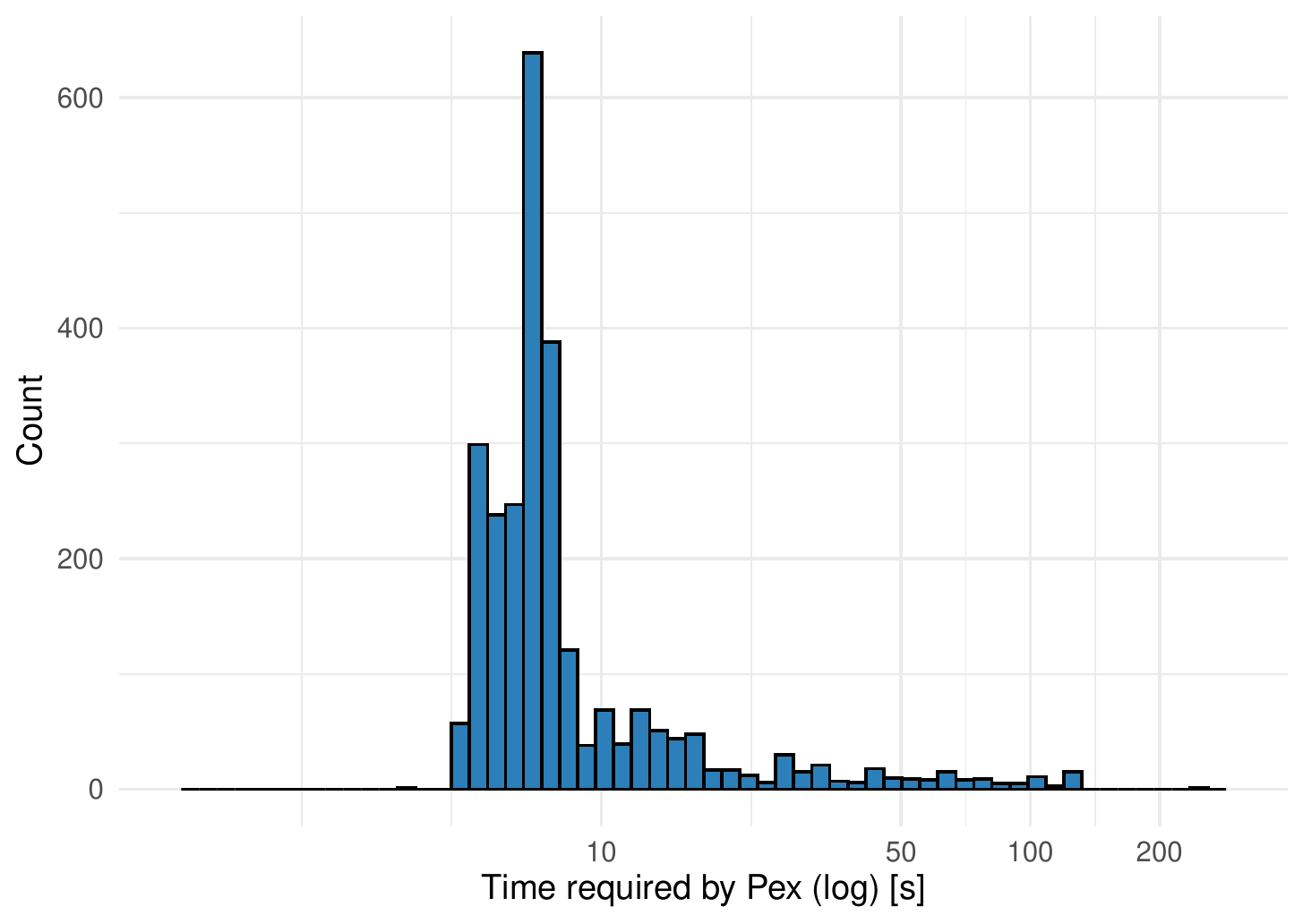}
    \caption{The distribution of time required for test generation by Pex without any isolation. Note the logarithmic scale.}
    \label{fig:pextime-pexonly}
\end{figure}

\begin{table}
    \centering
    \setlength{\tabcolsep}{3pt}
    \caption{The basic descriptive statistics of the test generation time in the 2596 executions performed by Pex without any isolation.}
    \label{tab:pextime-pexonly-stats}    
    \begin{tabular}{l c c c c c}\toprule 
        \emph{Variable} & \emph{Min} & \emph{Med} & \emph{Mean} & \emph{Max} & \emph{SD} \\
        \midrule
        TTestGeneration [s] & 3.65 & 7.06 & 11.54 & 237.31 & 162.20 \\
        \bottomrule
    \end{tabular}
\end{table}

In terms of time required by Pex, we show the basic descriptive statistics in Table~\ref{tab:pextime-pexonly-stats}. The median time required is 7.06 seconds and the mean is 11.54, but this contains those cases, where Pex did not generate any tests. The maximum time required was due to reaching some boundaries of the test generation. Figure~\ref{fig:pextime-pexonly} shows the distribution of the time required for test generation by Pex on a logarithmic scale. Note that the largest portion of the executions were performed under 10 seconds.

\subsubsection{RQ1: Code coverage improvement}

\begin{figure*}[!htb]
    \noindent\begin{minipage}{.47\textwidth}
        \centering
        \includegraphics[trim={0 0.8cm 0 0.8cm},width=\columnwidth]{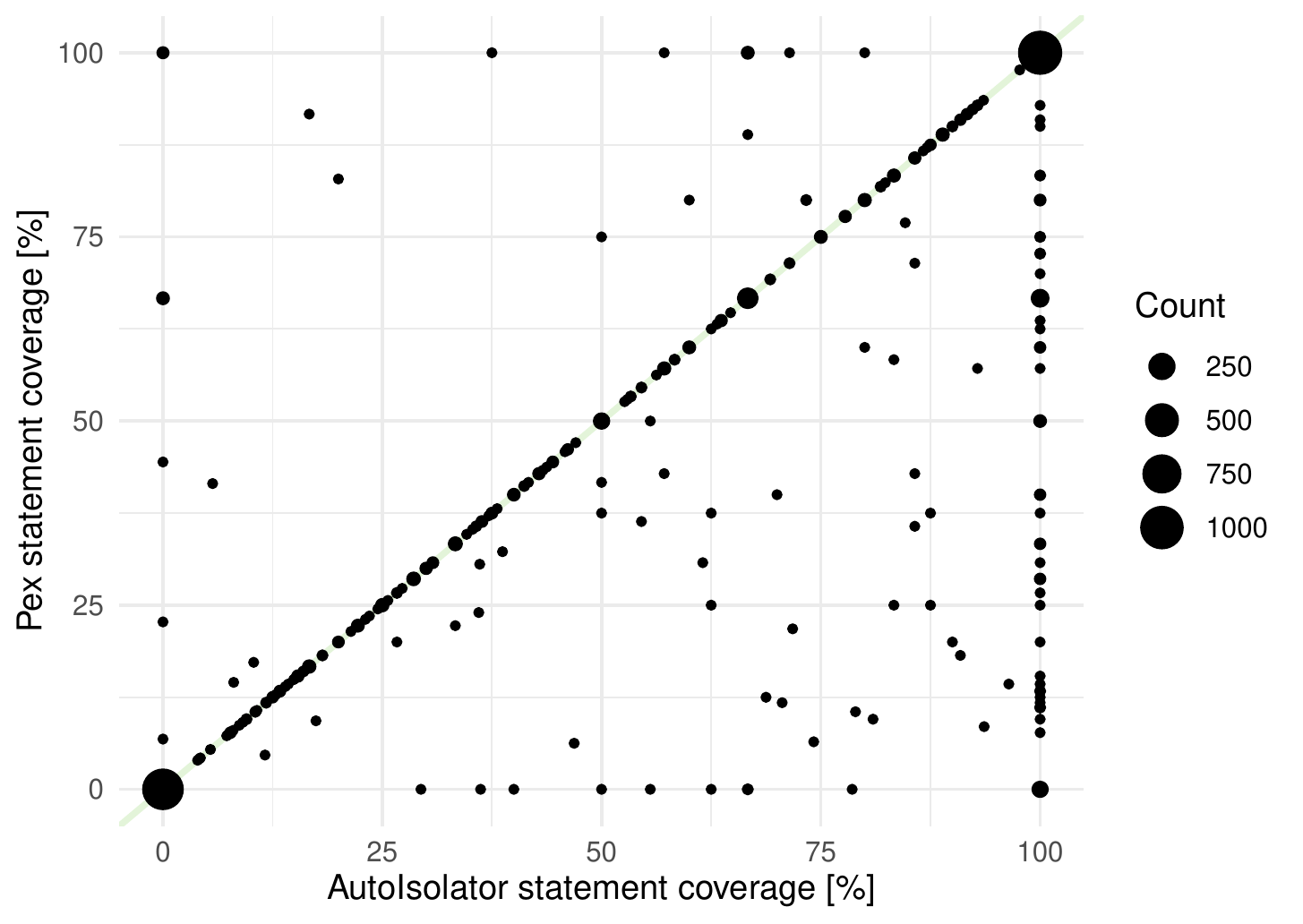}
        \caption{Comparison of statement coverages reached by Pex itself and with AutoIsolator.}
        \label{fig:scatter-sc}
    \end{minipage}\hfill
    \begin{minipage}{.47\textwidth}
        \centering
        \includegraphics[trim={0 0.8cm 0 0.8cm},width=\columnwidth]{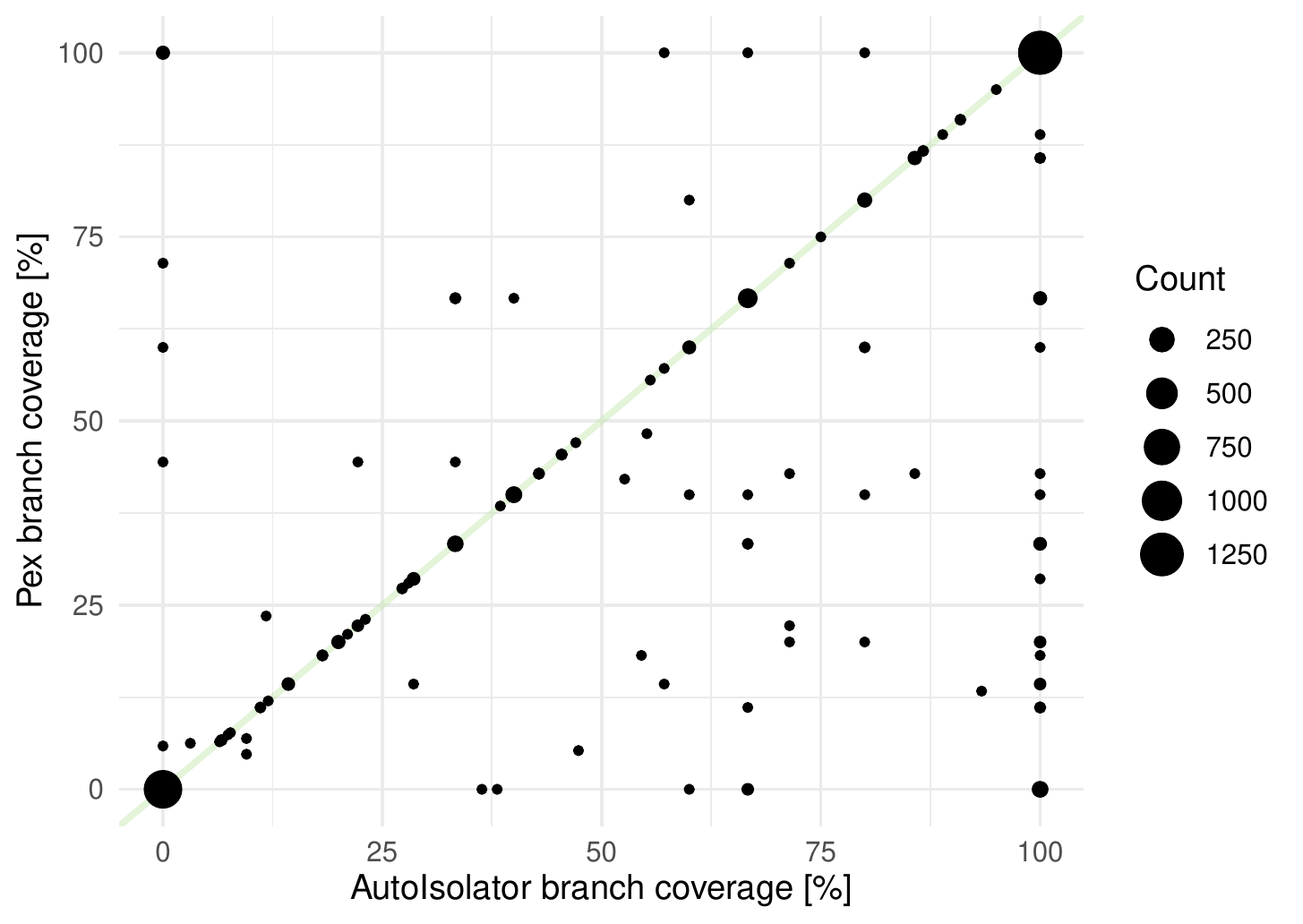}
        \caption{Comparison of branch coverages reached by Pex itself and with AutoIsolator.}
        \label{fig:scatter-bc}
    \end{minipage}
\end{figure*}

In this section, our goal is to discover, whether the automated isolation approach and the implemented tool can improve the coverage reached by generated white-box tests. First, we provide an overview in Figure~\ref{fig:scatter-sc} and Figure~\ref{fig:scatter-bc} about both types of coverage differences in the two observed cases: using Pex only, and using Pex with AutoIsolator. Out of the 2596 examined cases, there are 2361 where AutoIsolator did not affect the statement coverage (diagonal line in Figure~\ref{fig:scatter-sc}). In \emph{195 cases AutoIsolator could improve the statement coverage with an average of 52.95\% and median of 37.50\% additional coverage}. In terms of branch coverage, there were 2456 equal cases, while \emph{AutoIsolator could improve the branch coverage in 111 cases by an average of 67.58\% and median of 66.67\%}.

In the remaining 40 and 29 cases for statement and branch coverage, respectively, the execution with AutoIsolator reached a lower value. \emph{All of these} were caused by a known limitation of the implemented tool already mentioned in Sect.~\ref{sec:impl}: the tool does not support isolating transformations for operators applied on external (uninitialized) objects. Reaching a transformed code part, which contains this problematic structure yields an exception, which halts the execution. The most problematic cases were such that these operator usages were inside a constructor, and thus Pex could not instantiate the class under test.

Note that for statement and branch coverage there were 1059 and 1297 cases, respectively, in which Pex reached 100\% by itself, and thus AutoIsolator could not have improved the values in any ways.

\definecolor{darkpastelgreen}{rgb}{0.01, 0.75, 0.24}
\begin{table}
    \centering
    \setlength{\tabcolsep}{3pt}
    \caption{The basic descriptive statistics of the variables observed during the 2596 executions performed by Pex with automated isolation by AutoIsolator.}
    \label{tab:ai-basic-stats}    
    \begin{tabular}{l c c c c c}\toprule 
        \emph{Variable} & \emph{Min} & \emph{Med}  & \emph{Mean} & \emph{Max} & \emph{SD} \\
        \midrule
        TC & 0 & 1 & 2.79 & 114 & 5.89 \\
        SC [\%] & 0 & 66.67 & 56.08 (\textcolor{darkpastelgreen}{\textbf{+3.25}}) & 100 & 45.72 \\
        BC [\%] & 0 & 100.0 & 59.07 (\textcolor{darkpastelgreen}{\textbf{+2.21}}) & 100 & 46.52 \\
        \bottomrule
    \end{tabular}
\end{table}

Table~\ref{tab:ai-basic-stats} summarizes the basic descriptive statistics of the observed coverage values for the execution performed with AutoIsolator support. The number of generated tests slightly decreased, however it can be noticed that the statement and branch coverage values obviously improved (the mean gain compared to Pex-only execution is marked with green in the table).

\begin{figure}[!ht]
    \noindent\begin{minipage}{.47\columnwidth}
        \centering
        \includegraphics[trim={0 0.8cm 0 0.8cm},width=\columnwidth]{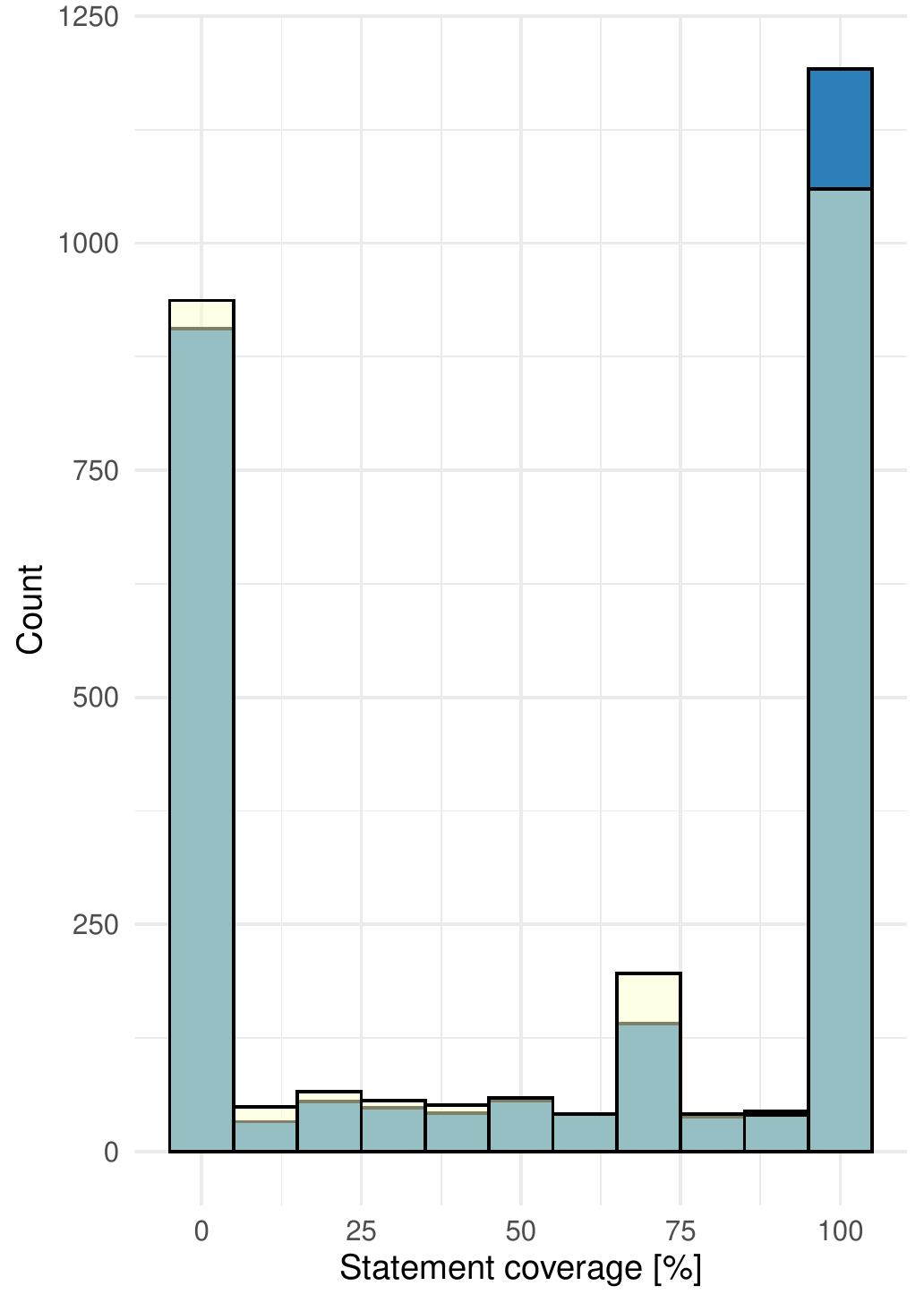}
        \caption{The distribution of statement coverage achieved by Pex with AutoIsolator (blue) compared to using Pex only (yellow).}
        \label{fig:ai-sc-hist}
    \end{minipage}\hfill
    \begin{minipage}{.47\columnwidth}
        \centering
        \includegraphics[trim={0 0.8cm 0 0.8cm},width=\columnwidth]{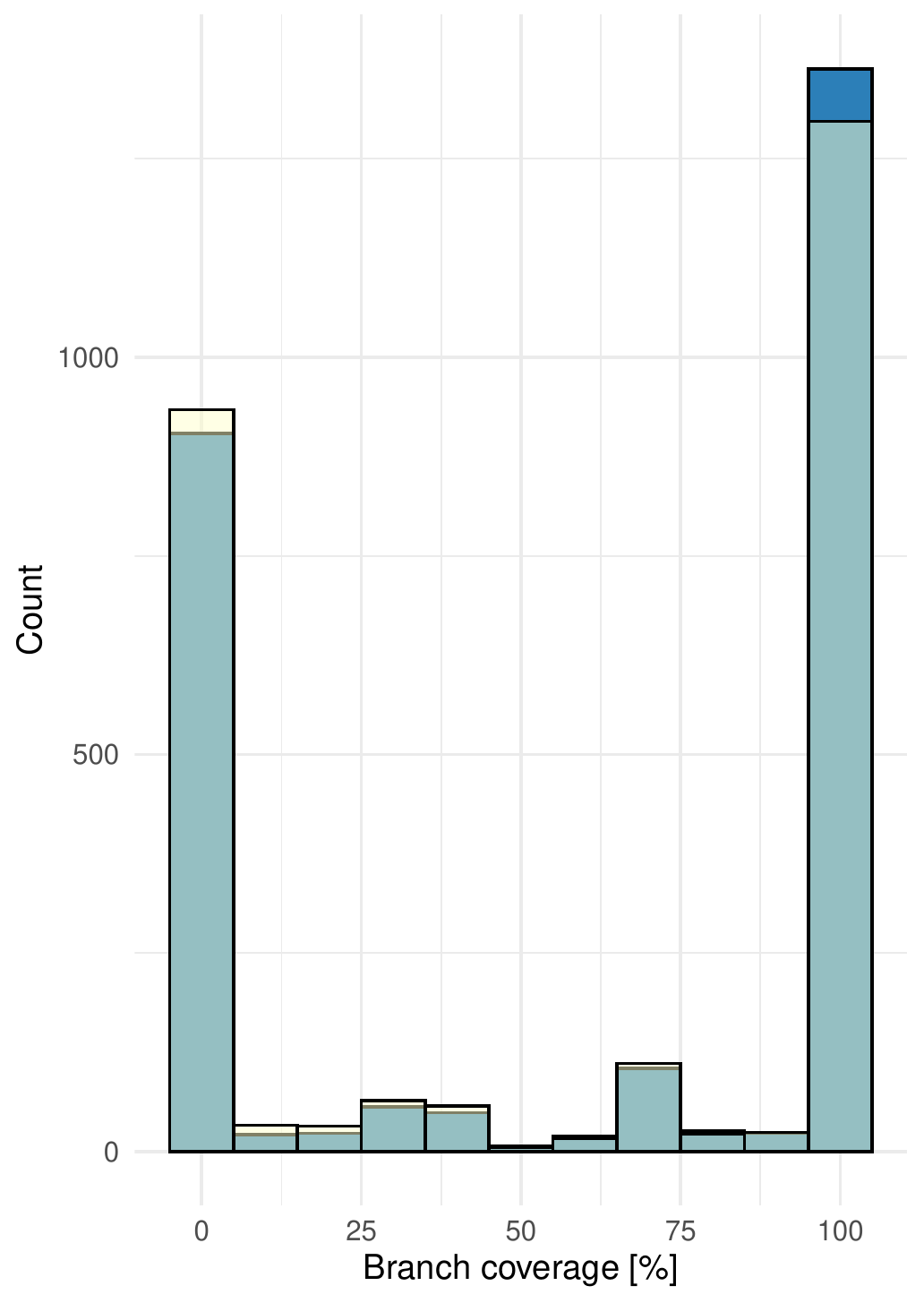}
        \caption{The distribution of branch coverage achieved by Pex with AutoIsolator (blue) compared to using Pex only (yellow).}
        \label{fig:ai-bc-hist}
    \end{minipage}
\end{figure}

Figure~\ref{fig:ai-sc-hist} and Figure~\ref{fig:ai-bc-hist} shows the distributions of statement and branch coverages, respectively, for the executions performed with the help of AutoIsolator (blue) compared to the case of using Pex only (yellow). It can be seen that both types of coverages indicate an increase in the number of cases having 100\% code coverage.

\medskip\noindent\textbf{Answer for RQ1.} Based on the results above, almost 200 of the analyzed 2596 methods were sensitive to isolation in terms of statement coverage (SC). For branch coverage (BC), more than 100 cases were subjects to coverage increase. Note that in cases where Pex reached 100\% by itself, there was no room for improvement to AutoIsolator. Nevertheless, the scale of the increase (52.95\% mean additional SC, and 67.58\% mean additional BC) indicate that if the automated isolation approach was able to help, then it was clearly effective in improving both types of code coverages.

\subsubsection{RQ2: Additional time required}

In RQ2, we investigate what is the additional time required by AutoIsolator in the whole test generation process. This is an important aspect, because white-box test generation aims to reduce the testing efforts invested into to the whole development process, yet if the automated isolation approach increases the time on a large extent, it may not worth using it. To find the answer, we measured the time spent by AutoIsolator in each of its steps and we compare them against the time required by Pex for test generation. In Table~\ref{tab:ai-time} we summarize the basic statistics of the results.

\begin{figure}[!ht]
    \centering
    \includegraphics[trim={0 0.8cm 0 0.8cm},width=\columnwidth]{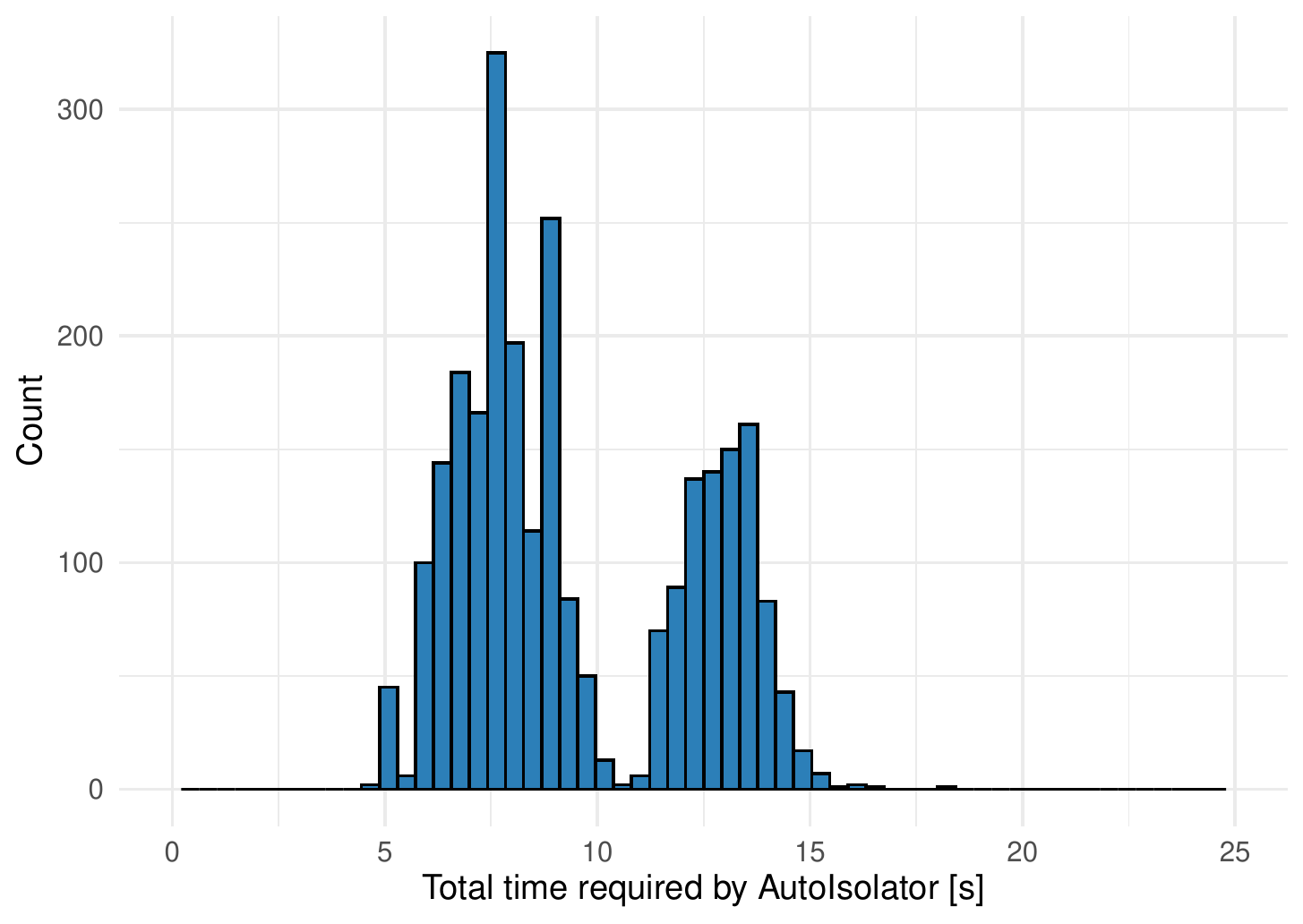}
    \caption{The distribution of the total time required for AutoIsolator itself (without test generation of Pex).}
    \label{fig:ai-time}
\end{figure}

AutoIsolator transformed the unit under tests in a median of 2.78 seconds (TTransformation), which includes the traversal and the possible replacements of the nodes in the abstract syntax trees. Note that this step also includes semantic type analyses as well. The code generation part (TCodeGeneration) -- as it is expected -- usually took a very short amount of time with a median of 0.4 seconds and a maximum of 2.73 seconds. Finally, the compilation part (TCompilation), which is intuitively depends on the size of the project, took 6.11 seconds on average with median of 5.43. The largest unit under test and its project took 16.66 seconds to compile. Summing up the previous three values yields that the time required by AutoIsolator is 9.54 seconds on average (having a median of 8.72). The longest execution for AutoIsolator was 26.14 seconds. The distribution of these values are shown in Figure~\ref{fig:ai-time}. Comparing these values to the times Pex spent with test generation indicates that the time of AutoIsolator moves on a very similar scale as the time required by Pex itself with (TTestGeneration in Table~\ref{tab:ai-basic-stats}) or without (TTestgeneration in Table~\ref{tab:pextime-pexonly-stats}) isolation. 

\begin{table}
    \centering
    \setlength{\tabcolsep}{3pt}
    \caption{The basic descriptive statistics of the required times in the 2596 executions performed by AutoIsolator, and Pex with isolation.}
    \label{tab:ai-time}    
    \begin{tabular}{l c c c c c}\toprule 
        \emph{Variable} & \emph{Min} & \emph{Med} & \emph{Mean} & \emph{Max} & \emph{SD} \\
        \midrule
        TTransformation [s] & 1.00 & 2.78 & 3.301 & 9.02 & 1.05 \\
        TCodeGeneration [s] & 0.31 & 0.40 & 0.42 & 2.73 & 0.10 \\
        TCompilation [s]    & 2.68 & 5.43 & 6.11 & 16.66 & 1.80 \\
        \midrule
        AutoIsolator Total [s]  & 4.81 & 8.72 & 9.54 & 26.14 & 2.76 \\ 
        TTestGeneration [s]     & 3.76 & 7.03 & 9.68 & 129.42 & 11.83 \\
        \bottomrule
    \end{tabular}
\end{table}

\medskip\noindent\textbf{Answer for RQ2.} Considering the results mentioned in the section, AutoIsolator usually doubles the time required for white-box test generation with Microsoft Pex as the values are very similar to what Pex requires by itself -- except the maximum time. These values however does not seem to be a large gambit, because they are still some dozens of seconds in total, yet the coverage achieved may be a much larger as considered in RQ1.

\subsection{Discussion}

\subsubsection{False positives}

\begin{figure}[!ht]
    \centering
    \includegraphics[trim={0 0 0 0},width=\columnwidth]{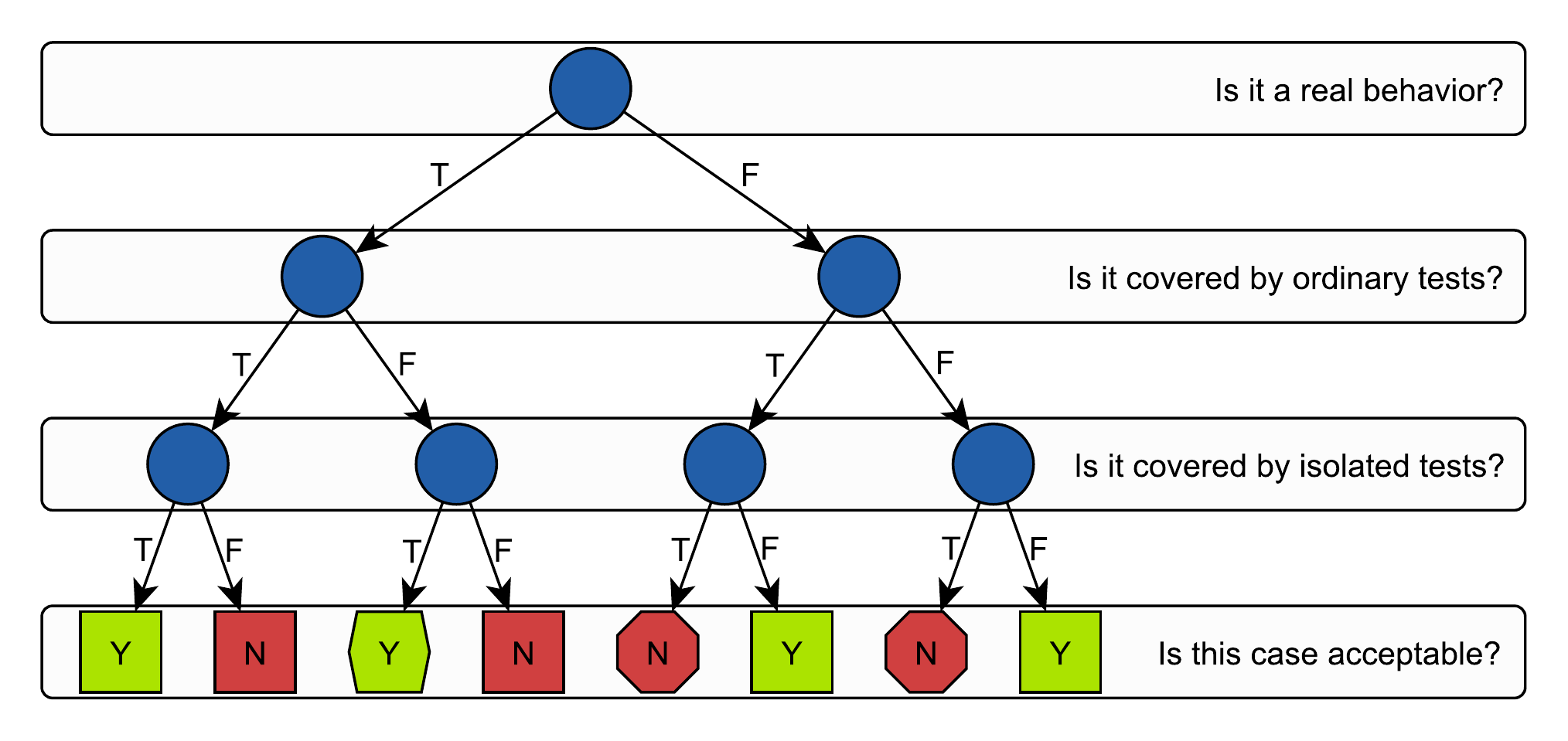}
    \caption{A decision tree providing an overview of the false positive problem with isolated isolated tests.}
    \label{fig:behavior}
\end{figure}

Our transformation approach automatically isolates all dependencies of a selected code unit and lets the white-box test generator to put behavior into the fake methods. This process yields that the white-box test generator can inject arbitrary values into the unit, which could not be valid for real executions (\emph{false behavior}). See Figure~\ref{fig:behavior} for an overview of this concern described with a decision tree. The tree can be used to decide whether a given combination of a) the behavior being checked, b) the original tests' coverage, and c) the isolated tests' coverage is a good (acceptable for practical use) or a bad case.

The real behavior induced by the original code is the starting point of the whole automated isolation process. If the external invocations inside this code are replaced with fake ones, then the number of possible behaviors is likely to increase. For instance, the sandbox can return a value that otherwise not possible in the original dependency (e.g., due to internal checks). An ordinary test suite on the original code may not cover all real behavior of the original code. For example, some errors are really hard to trigger in the external dependencies without isolation (e.g., permission issues in the file system).

If the tests use isolation, they can cover additional behavior not reachable from ordinary tests. There are two consequences of this: on one hand, they may cover otherwise uncoverable behaviors of the original code (this is the benefit of using the parameterized sandbox -- marked with a hexagon leaf in the decision tree), on the other hand, they may introduce tests triggering false behavior (marked as an octagon in the leaves of the decision tree). 

We hereby propose two possible solutions to tackle false behavior: 1) automatically extracted restrictions, 2) user-defined restrictions. For automatically extracting behavioral restrictions, one could use symbolic execution on the given dependency to extract procedural summaries like in compositional symbolic execution \cite{demand-driven}. For the latter, we already proposed an approach by which users can restrict the isolated behavior for test generation using an intuitive description of behaviors \cite{user-defined}.

\subsubsection{Practical applicability}

Using our approach, the white-box tests are generated on the transformed code. This implicates that -- most of the time -- these tests are a) not directly executable on the original, untouched code, and b) might not trigger the same behavior as in isolation. To tackle this issue, we hereby define an approach by which such tests remain applicable in practice.

The execution of tests that were generated on the transformed, isolated code requires the transformed code itself. This would require a fully separate handling of non-isolated and isolated source including their versioning as well. The concept of \emph{isolated tests} introduces a new type of tests for the test execution frameworks: when such type of test is being executed, then the framework -- with the help of an extension -- automatically transforms the code under test prior to the test execution. This way the execution of the isolated tests will remain transparent to the user and can be handled together with other types of test, such as regression tests. Most of the popular TEFs (e.g., NUnit, MSTest, JUnit) support these kinds of extensions.


\section{Conclusions and future work}\label{sec:conclusions}

This paper presented a novel approach and its evaluation on supporting automated white-box test generation via automated isolation to overcome the difficulties caused by external dependencies. The approach uses abstract syntax tree transformations to replace invocations and other accesses calling outside of the unit previously defined. This approach tackles most of the drawbacks of other approaches like the combination of white-box test generation with concrete mocking frameworks. We implemented the approach in a ready-to-use tool (AutoIsolator) for enhancing the test generation process of Microsoft Pex -- one of the most advanced white-box test generators.

To check whether AutoIsolator really does alleviate white-box test generation, we designed an experiment to quantify the improvements in terms of statement and branch coverage. Meanwhile, we also measured the time required by AutoIsolator to make sure that the transformations and the related steps do not take unreasonable amount of time compared to the test generation itself. The evaluation was performed on 10 randomly selected open-source C\# projects from GitHub. We generated tests with Pex against 2596 methods that have more than 38.000 lines of code: once with, and then without the support of AutoIsolator. The results not only serve as a baseline for our technique, but stands out as a unique overview of the performance of Pex on open-source projects.

The results imply that AutoIsolator is able to increase both statement and branch coverage in given cases by ~50\% on average. However, in almost one third of the cases, Pex was not able to generate any tests (even with the help of AutoIsolator), which might indicate that external dependencies might not be the main issue of white-box test generators. Also, it is important to note that AutoIsolator requires an average of 5-10 seconds to complete its full transformation process, which is on the same order of magnitude as the time of the test generation process.

We would like to enhance the approach with a sophisticated algorithm on how to handle the states of the externally-typed objects used inside the unit under test. Also, we would like to extend the support of the C\# grammar's corner cases to increase the external validity of the evaluation (these were skipped for now). Finally, a user-oriented experiment (e.g., a think-aloud study) may be required to grasp knowledge about the practical applicability from a user's perspective.
\bibliography{references}




\end{document}